\documentclass[acmsmall,screen]{acmart}

\usepackage[utf8]{inputenc}

\usepackage{amsfonts}
\usepackage{makecell,multirow}
\usepackage{tcolorbox}
\usepackage{xcolor}
\usepackage{listings}
\usepackage{caption,subcaption}
\usepackage{threeparttable}
\usepackage{bbding}
\usepackage{graphicx}
\usepackage{textcomp}
\usepackage{paralist}
\usepackage{fancyhdr}
\usepackage{booktabs}
\usepackage{url}
\usepackage{pdflscape}
\usepackage{float}
\usepackage{enumitem}
\usepackage{newunicodechar}

\newunicodechar{₁}{$_1$}
\newunicodechar{≤}{\leq{}}

\definecolor{dkgreen}{rgb}{0,0.6,0}
\definecolor{gray}{rgb}{0.4,0.4,0.4}
\definecolor{mauve}{rgb}{0.58,0,0.82}
\definecolor{darkblue}{rgb}{0.0,0.0,0.6}
\definecolor{lightblue}{rgb}{0.0,0.0,0.9}
\definecolor{cyan}{rgb}{0.0,0.6,0.6}
\definecolor{darkred}{rgb}{0.6,0.0,0.0}
\definecolor{yellow}{RGB}{255,255,153}
\definecolor{grey}{RGB}{220,220,220}
\definecolor{green}{RGB}{0,100,0}

\definecolor{KWColor}{rgb}{0.37,0.08,0.25}
\definecolor{CommentColor}{rgb}{0.133,0.545,0.133}
\definecolor{StringColor}{rgb}{0,0.126,0.941}
\definecolor{commentgreen}{RGB}{2,112,10}
\definecolor{eminence}{RGB}{108,48,130}
\definecolor{weborange}{RGB}{255,165,0}
\definecolor{frenchplum}{RGB}{129,20,83}

\lstdefinelanguage{json}{
  basicstyle=\ttfamily\footnotesize,
  showstringspaces=false,
  breaklines=true,
  frame=lines,
  literate=
   *{0}{{{\color{blue}0}}}{1}
    {1}{{{\color{blue}1}}}{1}
    {2}{{{\color{blue}2}}}{1}
    {3}{{{\color{blue}3}}}{1}
    {4}{{{\color{blue}4}}}{1}
    {5}{{{\color{blue}5}}}{1}
    {6}{{{\color{blue}6}}}{1}
    {7}{{{\color{blue}7}}}{1}
    {8}{{{\color{blue}8}}}{1}
    {9}{{{\color{blue}9}}}{1}
    {:}{{{\color{red}:}}}{1}
    {,}{{{\color{red},}}}{1}
    {"}{{{\color{orange}"}}}{1}
}

\lstdefinestyle{codeblock}{
  basicstyle=\ttfamily\footnotesize,
  breaklines=true,
  breakatwhitespace=false,
  columns=fullflexible,
  postbreak=\mbox{\textcolor{gray}{$\hookrightarrow$}\space},
  showstringspaces=false,
  frame=single,
  numbers=left,
  numberstyle=\tiny\color{gray},
  numbersep=8pt,
  abovecaptionskip=5pt,
  belowskip=1em,
  captionpos=b,
}

\AtBeginDocument{%
  \providecommand\BibTeX{{%
    \normalfont B\kern-0.5em{\scshape i\kern-0.25em b}\kern-0.8em\TeX}}}

\setcopyright{acmcopyright}
\copyrightyear{2025}
\acmYear{2025}
\acmDOI{XXXXXXX.XXXXXXX}

\acmJournal{TOSEM}
\acmVolume{X}
\acmNumber{Y}
\acmArticle{1}
\acmMonth{7}

\begin{document}
\title{GDPR-Bench-Android: A Benchmark for Evaluating Automated GDPR Compliance Detection in Android}

\author[H Ran]{Huaijin Ran}
\email{seventeen17510@gmail.com}
\authornotemark[1]
\affiliation{%
  \institution{Xi'an Jiaotong-Liverpool University}
  \city{SuZhou}           
  \country{China}
}
\author[H Zhang]{Haoyi Zhang}
\email{hyeliozhang@gmail.com}
\authornote{Co-first authors who contributed equally to this work.}
\affiliation{%
  \institution{Xi'an Jiaotong-Liverpool University}
  \city{SuZhou}           
  \country{China}
}
\author[X Tang]{Xunzhu Tang}
\authornote{Xunzhu Tang is the corresponding author (xunzhu.tang@uni.lu).}
\email{xunzhu.tang@uni.lu}
\affiliation{%
  \institution{University of Luxembourg}
  \city{Luxembourg}     
  \country{Luxembourg}
}

\begin{abstract}
Automating the detection of EU General Data Protection Regulation (GDPR) violations in source code is a critical but underexplored challenge. We introduce \textbf{GDPR-Bench-Android}, the first comprehensive benchmark for evaluating diverse automated methods for GDPR compliance detection in Android applications. It contains \textbf{1951} manually annotated violation instances from \textbf{15} open-source repositories, covering 23 GDPR articles at file-, module-, and line-level granularities. To enable a multi-paradigm evaluation, we contribute \textbf{Formal-AST}, a novel, source-code-native formal method that serves as a deterministic baseline. We define two tasks: (1) \emph{multi-granularity violation localization}, evaluated via Accuracy@\textit{k}; and (2) \emph{snippet-level multi-label classification}, assessed by macro-F1 and other classification metrics. We benchmark 11 methods, including eight state-of-the-art LLMs, our Formal-AST analyzer, a retrieval-augmented (RAG) method, and an agentic (ReAct) method. Our findings reveal that no single paradigm excels across all tasks. For Task 1, the ReAct agent achieves the highest file-level Accuracy@1 (17.38\%), while the Qwen2.5-72B LLM leads at the line level (61.60\%), in stark contrast to the Formal-AST method's 1.86\%. For the difficult multi-label Task 2, the Claude-Sonnet-4.5 LLM achieves the best Macro-F1 (5.75\%), while the RAG method yields the highest Macro-Precision (7.10\%). These results highlight the task-dependent strengths of different automated approaches and underscore the value of our benchmark in diagnosing their capabilities. All resources are available at: \url{https://github.com/Haoyi-Zhang/GDPR-Bench-Android}.
\end{abstract}

\maketitle




\section{Introduction}
\label{sec:introduction}

Modern software systems, particularly Android applications, routinely collect, process, and share vast amounts of personal data. This pervasive handling of user information exposes organizations to increasingly stringent regulatory requirements, particularly under the European Union's General Data Protection Regulation (GDPR)~\cite{voigt2017eu,albrecht2016gdpr}. Since its enactment in 2018, GDPR has established a landmark framework for privacy protection, introducing principles such as lawfulness, transparency, purpose limitation, and data minimization—with non-compliance penalties reaching up to 4\% of global annual turnover. For software engineers, translating these abstract legal mandates into concrete code constructs represents a formidable challenge, requiring them to bridge a fundamental semantic gap between high-level regulatory language and low-level implementation details.

Traditional approaches to ensuring compliance have significant limitations. Manual audits, while thorough, are labor-intensive, error-prone, and lack scalability in the face of modern development velocities and codebases~\cite{reardon2019fifty}. While established static analysis tools like FlowDroid~\cite{arzt2014flowdroid}, IccTA~\cite{li2015iccta}, and Amandroid~\cite{wei2014amandroid} offer automation for Android privacy analysis, they are designed to operate on compiled Android packages (.apk files) rather than directly on source code. This makes them incompatible with a source-code-native benchmark and less suitable for integration into early-stage development workflows. Furthermore, these tools, often tailored for security vulnerabilities, can generate excessive false positives and struggle to capture the holistic, context-dependent reasoning required for many GDPR provisions~\cite{bodden2013taming,christakis2016developers}. The challenge of testing privacy protection in complex software systems remains a fundamental research problem~\cite{basciftci2021privguard,degeling2018we}. This highlights the need for a source-code-centric formal baseline to enable fair comparisons against modern code-aware models.

Recent advances in large language models (LLMs) have demonstrated remarkable capabilities in natural language understanding and code generation~\cite{chen2021evaluating,wang2023codet5}, offering a promising new paradigm for automated code review and analysis~\cite{jimenez2024swebenchlanguagemodelsresolve}. Their capacity to jointly reason over both natural language (legal requirements) and code (implementation patterns) suggests strong potential for bridging the semantic gap in regulatory compliance. However, the ability of LLMs to rigorously identify GDPR violations—where nuanced interpretation of code semantics in relation to specific legal articles is essential—remains largely unexamined. Furthermore, a singular focus on LLMs overlooks other promising automated approaches, such as retrieval-augmented generation (RAG) and agentic methods, which may offer enhanced performance in long-context reasoning. Crucially, there exists no publicly available benchmark to systematically evaluate and compare the performance of these diverse automated methods on GDPR compliance detection directly from Android source code~\cite{jimenez2024swebenchlanguagemodelsresolve}.

To bridge this gap, we introduce \textbf{GDPR-Bench-Android}, the first comprehensive benchmark for assessing \textit{automated} GDPR compliance detection methods within the Android ecosystem. Our benchmark is built upon a substantially expanded dataset, now comprising over 1900 manually annotated violation instances drawn from fifteen diverse open-source Android applications. To establish a robust non-LLM baseline, we designed and implemented \textbf{Formal-AST}, a novel formal method tailored for source-code analysis. This expanded corpus covers a wide range of GDPR articles, with each violation meticulously labeled at file-, module-, and line-level granularities. Based on this dataset, we formulate two complementary evaluation tasks:

\begin{enumerate}
  \item \emph{Multi-granularity violation localization}, where a method must rank the most likely violated GDPR articles at three code scopes (file, module, and line), evaluated using Accuracy@\,$k$ metrics;
  \item \emph{Snippet-level multi-label classification}, where a method must assign all applicable articles to individual code fragments, assessed by exact-match accuracy and macro-averaged precision, recall, and F1 scores.
\end{enumerate}

\begin{figure}[tbp]
    \centering
    \includegraphics[width=\textwidth]{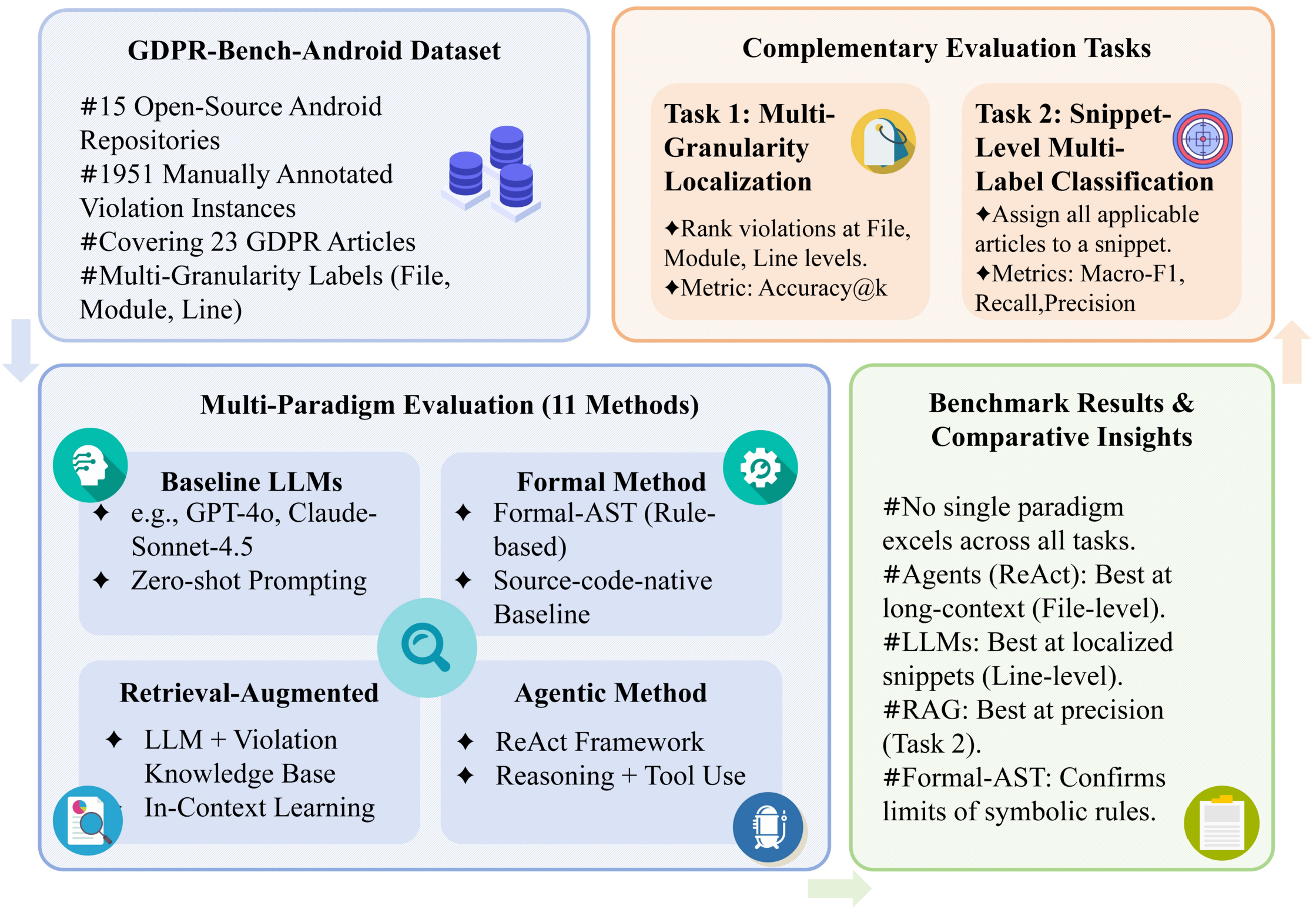}
    \caption{A High-Level Overview of the GDPR-Bench-Android Framework.}
    \label{fig:overview1}
\end{figure}

We conduct a comprehensive evaluation of eleven methods, including eight state-of-the-art LLMs (e.g., GPT-4o, Claude-Sonnet-4.5), our novel formal static analysis method (Formal-AST), a retrieval-augmented generation (RAG) approach~\cite{lewis2020retrieval}, and an agentic (ReAct) approach~\cite{yao2023react}. Our empirical results reveal key insights into the strengths and weaknesses of different paradigms. For instance, while LLMs excel at pinpointing violations in narrow contexts (line-level Accuracy@1 up to 61.60\%), agentic methods show superior performance in holistic, long-context reasoning at the file level. This multifaceted evaluation underscores the necessity of a diverse benchmark to guide the development of robust and reliable automated compliance tools. A high-level summary of our updated benchmark design is illustrated in Figure~\ref{fig:overview1}.

The contributions of this paper are fivefold:
\begin{itemize}
    \item \textbf{GDPR-Bench-Android Dataset}: An expanded and curated collection of over 1900 GDPR violation instances from 15 Android repositories, with rich annotations at file, module, and line levels.
    \item \textbf{A Novel Formal Method}: We design and implement Formal-AST, a source-code-native, rule-based detection method. It serves as a crucial non-LLM baseline and is, to our knowledge, the first formal approach specifically tailored for GDPR compliance analysis directly on Android source code.
    \item \textbf{Multi-Paradigm Evaluation}: A comprehensive benchmark of 11 automated methods, including baseline LLMs, our formal static analyzer, a retrieval-augmented method, and an agentic method, providing the first comparative insights into their capabilities for GDPR compliance detection.
    \item \textbf{In-depth Empirical Analysis}: A detailed analysis of how different automated methods perform across various code granularities, highlighting the trade-offs between symbolic, neural, and agentic reasoning for legal compliance tasks.
    \item \textbf{Open Release}: All data, prompts, and evaluation scripts are publicly available at \url{https://github.com/Haoyi-Zhang/GDPR-Bench-Android} to foster further research in AI-driven privacy compliance assessment.
\end{itemize}

Our work provides a standardized benchmark for evaluating a wide range of automated methods on GDPR compliance detection. The empirical results reveal distinct performance characteristics, highlighting that while certain approaches excel at localized pattern matching, others are better suited for broad contextual reasoning. GDPR-Bench-Android offers a reproducible framework for measuring progress and enables a systematic comparison between different approaches to this challenging and critical task.

The remainder of this paper is organized as follows. Section~\ref{sec:background_motivation} provides essential background on GDPR. Section~\ref{sec:related} situates our work within the broader landscape of compliance checking approaches. Section~\ref{sec:experiments} details our experimental methodology. Section~\ref{subsec:results} presents quantitative results. Section~\ref{subsec:analysis} provides in-depth analysis and error diagnosis. Finally, Section~\ref{sec:conclusion} discusses future directions and concludes the paper.

\section{Background and Motivation}
\label{sec:background_motivation}

\subsection{The Compliance Burden of Data Protection Laws}

The GDPR, enacted in 2018, represents a landmark legislation aimed at ensuring lawful, transparent, and purpose-limited processing of personal data within the EU~\cite{voigt2017eu}. It introduces foundational principles such as data minimization, user consent, and accountability, with non-compliance penalties reaching up to 4\% of global annual turnover. While the law targets organizations broadly, its technical implications permeate the entire software development lifecycle (SDLC)~\cite{voigt2017eu}. For software engineers, translating these abstract legal mandates into concrete code constructs represents a formidable challenge. The obligation of "data protection by design and by default" (Article 25), for instance, requires privacy considerations to be integrated from the earliest stages of requirements engineering and system architecture~\cite{notario2015pripare, colesky2018privacy}, rather than being treated as an afterthought. However, studies have shown that developers face significant practical barriers in implementing GDPR principles throughout the software development lifecycle~\cite{alhazmi2021listening}. This necessitates a fundamental shift in development practices, where legal compliance becomes a primary engineering concern.

\subsection{The Semantic Gap Between Law and Source Code}

Despite the normative clarity of the GDPR, translating its high-level requirements into low-level software logic remains a fundamental challenge. This "semantic gap" manifests in several dimensions. First, there is a mismatch between legal intent and code implementation. A GDPR principle like "purpose limitation" is about the intended use of data, whereas a code function is about its technical manipulation; the same function might be compliant in one call context but non-compliant in another. Second, legal text governs "personal data," an abstract concept, while code operates on concrete data types like strings and integers. The determination of whether a variable like \textbf{user\_id} constitutes personal data depends on its context within the broader system.

For example, the simple act of retrieving a device identifier, as shown in Listing~\ref{lst:semantic_gap_example}, is contextually ambiguous. It could be a compliant part of device management for security purposes or an illicit act of user tracking without consent. This distinction is impossible to make from the code snippet alone and depends entirely on surrounding code, user-facing consent flows, and the overall system architecture~\cite{harkous2018polisis}. This contextual dependency renders simple pattern-matching insufficient for accurate compliance checking.

\begin{lstlisting}[caption={A simple API call illustrating the semantic gap. Its compliance depends on context not present in the snippet itself.}, label={lst:semantic_gap_example}, language=Java]
// Is this GDPR compliant? It depends.
String deviceId = telephonyManager.getDeviceId(); 
\end{lstlisting}

\subsection{Limitations of Traditional and Emerging Automated Approaches}

Traditional techniques for privacy auditing have long relied on static and dynamic analysis~\cite{livshits2005finding, yamaguchi2014modeling}. Tools like FlowDroid~\cite{arzt2014flowdroid}, Amandroid~\cite{wei2014amandroid}, and IccTA~\cite{li2015iccta} are powerful for taint analysis and detecting data leaks in Android applications. However, a critical limitation is that they are predominantly designed to operate on compiled Android packages (\textbf{.apk} files). This makes them unsuitable for a source-code-native benchmark like ours, which is crucial for identifying violations early in the development lifecycle (e.g., within CI/CD pipelines) before an application is even compiled. This architectural mismatch reveals a critical gap: the lack of a reproducible, deterministic formal baseline that operates directly on source code.

More recently, the advent of Large Language Models (LLMs) has introduced a promising new paradigm~\cite{chen2021evaluating, wang2023codet5, 10.1145/3749370}. Their ability to jointly reason over natural language (legal text, but also other artifacts like app reviews or issue reports) and source code offers potential for bridging the semantic gap~\cite{jimenez2024swebenchlanguagemodelsresolve, tang2024app}. Furthermore, advanced techniques such as RAG and agentic methods (e.g., ReAct) are emerging as powerful tools for handling tasks with long and complex contexts. However, while promising, the true capabilities and limitations of these modern AI-driven approaches for GDPR compliance at the source-code level remain largely unverified due to the absence of a standardized evaluation framework. This leaves critical research questions unanswered: How reliably can these models distinguish between compliant and non-compliant code that is semantically similar? Do they generalize across different programming languages and code styles? And how can their susceptibility to 'hallucinations' be managed in a high-stakes legal context where precision is paramount and a single false negative could lead to significant financial and reputational damage? Answering these questions requires a standardized evaluation framework.

\subsection{The Critical Need for GDPR-Bench-Android}

The dual limitations of existing approaches—the incompatibility of traditional APK-based tools with source-code analysis and the unevaluated nature of modern AI methods—create a pressing need for a new type of benchmark. This problem landscape and the motivational gap are illustrated conceptually in Figure~\ref{fig:motivation_gap}. Specifically, the field lacks three essential components for advancing automated GDPR compliance:

\begin{figure}[htbp]
    \centering
    \includegraphics[width=1\textwidth]{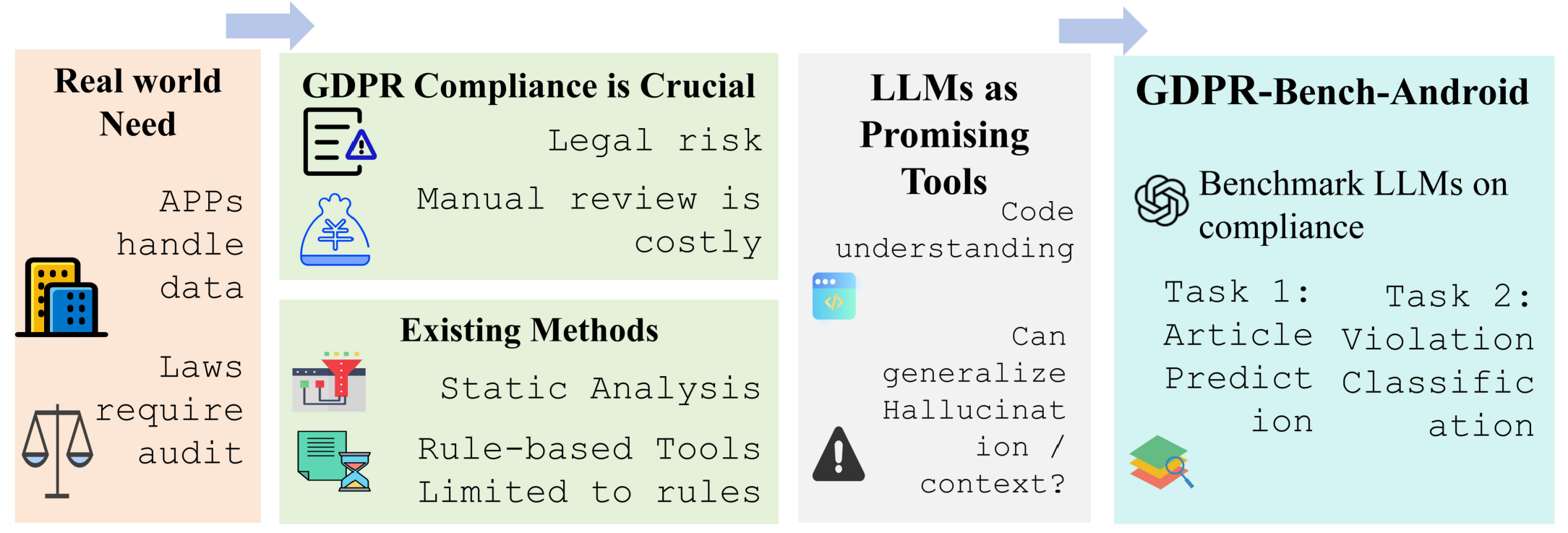}
    \caption{A conceptual overview of the problem motivation, illustrating the gap between the real-world need for GDPR compliance and the limitations of existing automated methods, which motivates the creation of our benchmark.}
    \label{fig:motivation_gap}
\end{figure}

\begin{enumerate}
    \item \textbf{A large-scale, legally-grounded source-code dataset.} While general-purpose code datasets like CodeXGLUE~\cite{lu2021codexglue} and Defects4J~\cite{just2014defects4j} exist, they focus on functional correctness or syntactic properties, lacking the legal context and specific violation patterns necessary for compliance tasks. This necessitates a specialized corpus.
    \item \textbf{Standardized tasks and metrics for fair comparison.} Without a common set of tasks and metrics, a method's reported performance remains isolated and incomparable, hindering the community's ability to track genuine progress and identify the most effective techniques for this domain.
    \item \textbf{A reproducible, source-code-native baseline.} The evaluation of opaque, non-deterministic models like LLMs requires a transparent and reproducible baseline to ground the analysis. The absence of a source-code native formal baseline makes it difficult to assess whether the complex reasoning of AI models provides a tangible benefit over traditional, deterministic logic.
\end{enumerate}

To address this multifaceted gap, we introduce \textbf{GDPR-Bench-Android}. Our work fills this void by delivering the three missing components. First, it provides a substantially expanded dataset of over 1900 manually annotated violation instances from fifteen real-world Android repositories. Second, recognizing the lack of a suitable baseline, we designed and implemented \textbf{Formal-AST}, a novel, source-code-native formal analysis method. This allows us, for the first time, to directly compare the performance of modern LLMs, agentic methods, and retrieval-augmented systems against a deterministic, rule-based approach on a common, open-source codebase. Finally, by providing a multi-faceted evaluation framework with well-defined tasks, our benchmark enables researchers to probe the nuanced capabilities of each method, from holistic reasoning on entire files to localized analysis of single code snippets.

\section{Related Work}
\label{sec:related}
This section situates \textbf{GDPR-Bench-Android} within the research landscape. We review traditional static analysis approaches for privacy, the application of LLMs in code analysis, the rise of agentic and retrieval-augmented methods in software engineering (SE), and existing benchmarks for code and legal compliance.

\subsection{Rule-Based and Static Analysis for Privacy}
\label{subsec:static_analysis}
Traditional automated compliance detection has heavily relied on static and dynamic taint-based analysis to uncover data misuse patterns, especially in Android. Foundational tools in this domain include \textbf{FlowDroid}~\cite{arzt2014flowdroid}, which performs precise data flow analysis; \textbf{IccTA}~\cite{li2015iccta}, which detects inter-component data leaks; and \textbf{Amandroid}~\cite{wei2014amandroid}, a general framework for security vetting. These tools are highly effective at tracking data flows from defined sources (e.g., location services) to sinks (e.g., network sockets)~\cite{reardon2019fifty}. Understanding Android's permission system is fundamental to these analyses~\cite{felt2011android}, and empirical studies have revealed critical security issues in real-world apps~\cite{chen2018mobile}. Additional approaches such as \textbf{TaintDroid}~\cite{enck2014taintdroid} have extended dynamic taint tracking to monitor runtime data flows in Android applications.

However, these traditional approaches face two major limitations in the context of our work. First, they struggle with the semantic ambiguity of legal requirements~\cite{voigt2017eu}. A data flow identified by \textbf{FlowDroid} is not inherently a GDPR violation; its legality depends on contextual factors like user consent or data minimization, which are difficult to encode in formal rules. Researchers have attempted to bridge this policy gap with tools like \textbf{Polisis}~\cite{harkous2018polisis} which analyzes natural language privacy policies, and PolicyLint~\cite{andow2016toward} which detects inconsistencies between policies and code. However, these methods often require extensive developer annotations or struggle with scalability~\cite{voigt2017eu}. Recent work on privacy policy analysis~\cite{zimmeck2019maps} has shown that even sophisticated automated tools face challenges in capturing the nuanced requirements of regulations like GDPR.

Second, and most critically for our evaluation, these established tools are overwhelmingly designed to operate on \textbf{compiled Android packages (\textbf{.apk} files)} or bytecode~\cite{arzt2014flowdroid, wei2014amandroid}. This architectural design, while powerful for post-deployment auditing, makes them fundamentally incompatible with a source-code-native benchmark intended for early-stage development workflows. This incompatibility creates a significant evaluation gap, as it prevents a direct "apples-to-apples" comparison between their symbolic reasoning and the semantic reasoning of source-code-native LLMs. Our work addresses this gap by introducing \textbf{Formal-AST}, a novel, deterministic, and multi-language analyzer that operates directly on Java and Kotlin source code. It serves as the first reproducible formal baseline specifically designed for this task, enabling a fair comparison between traditional symbolic logic and modern neural methods on a common source-code corpus.

\subsection{Large Language Models for Code Analysis and Security}
\label{subsec:llms_for_code}
The advent of Large Language Models (LLMs) has introduced a new paradigm for code understanding, synthesis, and repair~\cite{chen2021evaluating}. Models pretrained on massive code corpora through joint training on code and natural language, such as \textbf{CodeBERT}~\cite{feng2020codebert}, \textbf{CodeT5+}~\cite{wang2023codet5} and \textbf{StarCoder}~\cite{li2023starcoder}, exhibit strong semantic reasoning capabilities. In the security domain, LLMs show promise in identifying common vulnerabilities, though recent comprehensive evaluations reveal significant limitations in their detection capabilities~\cite{ding2024vulnerability}. Recent studies have also explored the application of LLMs to vulnerability detection and program repair~\cite{yu2023gptfuzzer}, demonstrating their potential in automated security analysis.

However, the application of LLMs to the nuanced field of legal compliance is nascent. While some studies have explored LLM performance on legal text~\cite{nay2024large}, their ability to map abstract legal articles (like GDPR) to concrete implementation flaws in source code is largely unexamined. \textbf{BigCodeBench}~\cite{bigcodebench2024} evaluates instruction-following on general code tasks but does not cover the specific domain of regulatory compliance. Our benchmark provides the first focused evaluation of baseline LLMs (e.g., GPT-4o, Claude Sonnet) on this specialized detection task.

\subsection{Agentic and Retrieval-Augmented Software Engineering}
\label{subsec:rag_and_agents}
Beyond standard zero-shot prompting, two advanced AI paradigms are emerging as powerful tools for complex SE tasks: retrieval-augmented generation (RAG) and agentic systems.

\paragraph{Retrieval-Augmented Generation} RAG enhances an LLM's generative capabilities by first retrieving relevant information from an external knowledge base. In the context of software engineering, this approach is gaining traction~\cite{gao2023retrievalaugmented}. Instead of relying solely on an LLM's pre-trained (and potentially outdated) knowledge, RAG can fetch context from project documentation, API guides, or—more recently—a repository of similar code snippets, known bugs, and security vulnerabilities. This retrieved context allows the LLM to perform more accurate, in-context learning, which has shown promise for tasks like vulnerability detection and code analysis. Our work evaluates a RAG method that operationalizes this concept by using our entire 1,900+ violation dataset as a specialized knowledge base, allowing the LLM to retrieve the most semantically similar compliant and non-compliant code examples before making a judgment.

\paragraph{Agentic Methods} Agentic methods, such as the ReAct framework~\cite{yao2023react}, empower LLMs to act as autonomous agents. These agents can iteratively reason, plan, and use a set of predefined tools to solve complex, multi-step problems. In SE, this paradigm is exemplified by the \textbf{SWE-bench} benchmark~\cite{jimenez2024swebenchlanguagemodelsresolve}, which evaluates agents on their ability to autonomously resolve real-world GitHub issues by using tools for file editing, searching, and test execution. Recent work on SWE-agent~\cite{jimenez2024swe} demonstrates that carefully designed agent-computer interfaces can significantly enhance automated software engineering capabilities. While agentic AI is being applied to tasks like debugging and automated coding, its application to the nuanced domain of legal compliance has not been explored. Our work provides the first evaluation of a ReAct-based agent specifically equipped with GDPR-centric tools (e.g., legal article lookup, code rule checking) and compares its long-context reasoning performance against other automated paradigms.

\subsection{Benchmarks for Code and Privacy Compliance}
\label{subsec:benchmarks}
Existing benchmarks for automated code analysis have historically focused on functional correctness rather than regulatory compliance. Foundational datasets like \textbf{Defects4J}~\cite{just2014defects4j} provide databases of known bugs to evaluate program repair. \textbf{CodeXGLUE}~\cite{lu2021codexglue} and \textbf{MBPP}~\cite{austin2021program} offer benchmarks for general code understanding and synthesis. While \textbf{SWE-bench}~\cite{jimenez2024swebenchlanguagemodelsresolve} presents realistic, long-horizon SE tasks, its focus remains on functional bug fixing, not legal compliance. More recent benchmarks such as \textbf{HumanEval}~\cite{chen2021evaluating} have further advanced code generation evaluation, but similarly do not address regulatory requirements.

Conversely, benchmarks in the legal-privacy domain have centered on \textit{textual artifacts}, not source code. \textbf{CLAUDETTE}~\cite{lippi2019claudette} and \textbf{Polisis}~\cite{harkous2018polisis} are seminal works for analyzing natural language privacy policies. Other datasets focus on analyzing privacy policies as textual artifacts rather than source code, rather than the underlying code that generates those logs.

\textbf{GDPR-Bench-Android} fills this critical and previously unaddressed gap. To our knowledge, it is the first benchmark to provide a large-scale (over 1,900 instances) corpus of manually annotated source-code violations grounded in specific GDPR articles. Furthermore, by including our novel \textbf{Formal-AST} baseline and evaluating RAG and ReAct methods, our work provides a unique and necessary framework for systematically comparing the efficacy of diverse automated paradigms—symbolic, neural, retrieval-based, and agentic—for the complex challenge of compliance checking directly within the software development lifecycle.

\section{Dataset}
\label{sec:dataset}

In this section, we detail the creation of GDPR-Bench-Android, explain its schema, and present a comprehensive statistical analysis. We structure our discussion into three parts: (1) \emph{Dataset Construction}, which describes our expanded data collection and human-in-the-loop annotation workflow; (2) \emph{Dataset Explanation}, which unpacks each field of our JSON schema and provides a representative example; and (3) \emph{Dataset Statistical Analysis}, which presents quantitative insights from our new, larger corpus.

\subsection{Dataset Construction}
\label{sec:dataset_construction}
The GDPR-Bench-Android corpus was assembled through a systematic, multi-phase annotation pipeline, designed to capture authentic violations of the European Union's GDPR in real-world Android applications. Our methodology balances computational pre-screening with expert oversight to ensure comprehensive coverage and high annotation quality.

\begin{figure}[htbp]
    \centering
    \includegraphics[width=\textwidth]{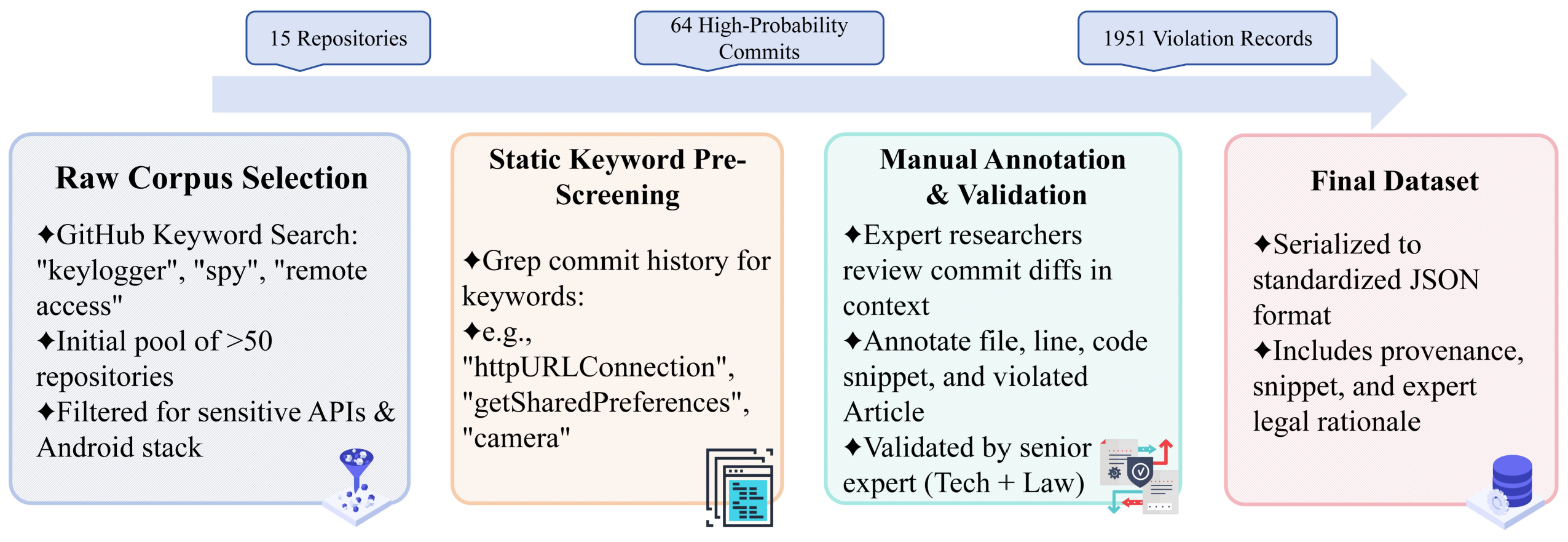}
    \caption{Updated Dataset Construction Pipeline.}
    \label{fig:pipeline}
\end{figure}

As illustrated in Figure~\ref{fig:pipeline}, the construction process unfolds across four consecutive phases.

\paragraph{Raw Corpus Selection}
We conducted a comprehensive survey of over fifty open-source Android projects hosted on GitHub. To identify this initial pool, we used privacy-related keywords (such as "keylogger", "spy", "remote access", "android permission"). We then applied a set of filtering criteria to this pool to select our final corpus. Our criteria required that: (a) the repository must handle user data via sensitive Android APIs (e.g., camera, location, contacts); and (b) it must employ technology stacks common to Android development (e.g., Kotlin, Java, or hybrid frameworks).

This process yielded our final dataset of \textbf{fifteen} repositories. The selection was strategic: we \emph{intentionally} included repositories identified as Remote Administration Tools (RATs), spyware (e.g., `AhMyth-Android-RAT`, `AndroSpy`), and other surveillance-focused applications. While these may not represent typical benign applications, they provide the richest source of clear-cut, unambiguous, and non-trivial GDPR violations, which is essential for establishing a robust ground truth for a benchmark. This is balanced with other general-purpose applications where violations are more subtle.

\paragraph{Static Keyword Pre-Screening}
To efficiently identify commits with a high likelihood of containing GDPR-relevant modifications, we employed a static keyword pre-screening approach instead of a "lightweight LLM classifier". We performed pattern-based searches across the repositories' commit histories for keywords related to sensitive data operations (e.g., `getSharedPreferences`, `httpURLConnection`, `write`, `log`, `camera`, `location`). This rule-based pre-filtering stage allowed us to narrow the manual review scope to 64 high-probability commits, significantly reducing the annotation effort. We focused on commits rather than code snapshots, as commit diffs often isolate specific functionality (e.g., "add feature X"), making it easier for annotators to pinpoint the introduction of a new violation in context.

\paragraph{Manual Annotation and Validation}
Expert annotators systematically examined each flagged commit within its complete repository context, identifying code fragments that breach one or more specific GDPR Articles. The annotation team consisted of three graduate researchers with expertise in software engineering and privacy. All final annotations were validated by a senior expert with a background in both technology and law to ensure legal accuracy. Our annotation guidelines established precise mappings to corresponding Article numbers (e.g., Article 6 for lawful basis, Article 32 for security). For each identified violation instance, annotators meticulously recorded:

\begin{itemize}[noitemsep]
    \item \textbf{app\_name}, \textbf{repo\_url}, and \textbf{commit\_id} to establish complete data provenance;
    \item \textbf{violated\_article} as a structured integer label corresponding to specific GDPR provisions;
    \item \textbf{code\_snippet\_path} specifying exact file location and line ranges, along with the corresponding \textbf{code\_snippet};
    \item a concise yet comprehensive \textbf{annotation\_note} articulating the specific legal rationale and violation context.
\end{itemize}

This intensive annotation phase yielded \textbf{1951} high-quality, non-duplicate violation instances. We acknowledge that our analysis is \textbf{static} and focuses on violations identifiable from source code. We do not perform run-time checks; therefore, a consent form present in a different, un-analyzed part of the codebase would not be captured. This is an inherent limitation of a static analysis benchmark, which we discuss further in Section~\ref{sec:conclusion}.

\paragraph{Quality Assurance and Consensus Building}
To ensure annotation reliability and consistency, each record underwent independent blind review by a second domain expert. Inter-annotator disagreements were systematically resolved through structured mediation sessions. All finalized annotation records were serialized into the standardized JSON format detailed in Table~\ref{tab:field_descriptions}. The complete construction pipeline for the expanded dataset required over 250 person-hours, balancing efficient pre-screening with rigorous human oversight.

\subsection{Dataset Explanation}
\label{sec:dataset_explanation}

The GDPR-Bench-Android corpus comprises 1951 meticulously annotated JSON-formatted records, each representing a distinct instance of non-compliance. Every record is systematically designed to capture both the technical implementation details and legal dimensions of a violation, thereby supporting a comprehensive spectrum of software engineering research tasks.

As detailed in Table~\ref{tab:field_descriptions}, our structured schema encompasses seven key fields that collectively provide a holistic view of each GDPR violation instance. This architecture ensures researchers can conduct multi-faceted analyses while maintaining traceability.

\paragraph{Provenance Metadata}
The foundational triad of fields (\textbf{app\_name}, \textbf{repo\_url}, and \textbf{commit\_id}) provides indispensable provenance information. By establishing direct linkage to specific GitHub repositories and commit SHA identifiers, we enable exact reconstruction of the software state in which each breach occurred. This traceability facilitates reproducibility, temporal analysis, and cross-project comparisons.

\paragraph{Legal Annotation Framework}
The \textbf{violated\_article} field encodes the precise GDPR Article infringed by a given snippet. Our annotation methodology allows for multi-labeling at the code-snippet level. A single code fragment can be associated with \emph{multiple} records in our dataset, each representing a distinct violation. For example, a single API call (as shown in Listing~\ref{lst:example_json}) might have one record citing a violation of Article 6 (Lawfulness of processing) and a separate record citing a violation of Article 32 (Security of processing) for the same line of code. This one-to-many relationship reflects the interconnected nature of GDPR obligations and allows our benchmark to rigorously test multi-label classification.

\paragraph{Code Context and Localization}
Precise localization of regulatory violations is achieved through the strategically paired fields \textbf{code\_snippet\_path} and \textbf{code\_snippet}. The \textbf{code\_snippet\_path} field provides granular specification of the exact file location and line range, while the \textbf{code\_snippet} field contains the corresponding raw source code lines that instantiate the regulatory breach. This explicit coupling supports advanced research methodologies that leverage both code structure and semantic information for compliance analysis.

\paragraph{Expert-Authored Legal Explanations}
The \textbf{annotation\_note} field provides concise, expert-authored rationales that establish explicit connections between identified code fragments and their corresponding legal infractions. Drawing inspiration from established legal annotation datasets such as CLAUDETTE~\cite{lippi2019claudette}, these justifications serve as authoritative gold-standard references for evaluating the quality and accuracy of model-generated legal explanations.

\begin{table}[H]
 \centering
 \caption{GDPR-Bench-Android JSON schema fields and their semantic roles}
 \label{tab:field_descriptions}
 \begin{tabular}{@{}ll@{}}
    \toprule
    \textbf{Field} & \textbf{Semantic Role} \\
    \midrule
    \textbf{app\_name} & Repository identifier enabling cross-project analysis \\
    \textbf{repo\_url} & URL facilitating contextual inspection and reproducibility \\
    \textbf{commit\_id} & Git SHA ensuring precise temporal provenance \\
    \textbf{violated\_article} & Integer label for the specific GDPR Article breached \\
    \textbf{code\_snippet\_path} & File path and line range for fine-grained localization \\
    \textbf{code\_snippet} & Source code fragment illustrating the regulatory breach \\
    \textbf{annotation\_note} & Expert rationale serving as ground truth for explainable AI \\
    \bottomrule
 \end{tabular}
\end{table}

To exemplify this structure, Listing~\ref{lst:example_json} presents two representative records from our new dataset. Both records point to the \emph{exact same line of code} in \textbf{MainActivity2.java}, but they are annotated with different violations. The first record flags the lack of a lawful basis (Article 6), while the second flags the lack of security measures (Article 32) for the same action. This multi-record annotation for a single snippet is what enables our multi-label classification task.

\begin{lstlisting}[
 % language=json,  % 
  caption={Representative GDPR-Bench-Android records illustrating the one-to-many, multi-label annotation structure for a single code snippet. Both records refer to the same line of code but cite different violations.},
  label={lst:example_json},
  basicstyle=\ttfamily\scriptsize, %
  breaklines=true,
  breakatwhitespace=false,
  columns=fullflexible,
  postbreak=\mbox{$\hookrightarrow$\space}, 
  frame=single,
  captionpos=b,
  aboveskip=0.5em,
  belowskip=0.5em
]
{
  "app_name": "Android_Spy_App",
  "repo_url": "https://github.com/abhinavsuthar/Android_Spy_App",
  "Commit_ID": "d524cef38b5526861a724f3f9b08b8b9a4a3d06a",
  "violated_article": 6,
  "code_snippet_path": "app/src/main/java/me/hawkshaw/test/MainActivity2.java: line 202",
  "code_snippet": "            manager.openCamera(camerId, stateCallback, null);\n",
  "annotation_note": "This code accesses the camera without verifying a lawful basis for processing personal data, violating GDPR Article 6."
},
{
  "app_name": "Android_Spy_App",
  "repo_url": "https://github.com/abhinavsuthar/Android_Spy_App",
  "Commit_ID": "d524cef38b5526861a724f3f9b08b8b9a4a3d06a",
  "violated_article": 32,
  "code_snippet_path": "app/src/main/java/me/hawkshaw/test/MainActivity2.java: line 202",
  "code_snippet": "            manager.openCamera(camerId, stateCallback, null);\n",
  "annotation_note": "This code does not document the processing activities related to camera access, violating GDPR Article 32 on security measures."
}
\end{lstlisting}

To further clarify these violations from a technical perspective for software engineering researchers, this single line of code exemplifies two distinct types of compliance failures:
\begin{itemize}[noitemsep]
    \item \textbf{Violation of Article 6 (Lawfulness of processing):} The \textbf{annotation\_note} (based on the provided JSON data) states the code "accesses the camera without verifying a lawful basis". For an engineer, this translates to a common implementation flaw: a sensitive API call (\textbf{manager.openCamera(...)}) is made without a preceding check to ensure user consent has been granted (e.g., a missing \textbf{if (userHasGivenConsent())} check or a failure to verify the result of a permission request).
    \item \textbf{Violation of Article 32 (Security of processing):} The note for the second record indicates the code "does not document the processing activities related to camera access". From a technical standpoint, this points to a failure in implementing "data protection by design" (part of Art. 25 and 32). The code lacks appropriate technical safeguards for handling the data, such as logging the access for auditing, ensuring the captured image data is encrypted at rest, or securely limiting who can access the data.
\end{itemize}

This structured approach enables three fundamental categories of research analysis:

\begin{enumerate}[noitemsep]
    \item \emph{Reproducible Localization and Verification}: Researchers can check out the exact commit and examine the source lines to reproduce or extend our annotations.
    \item \emph{Multi-label Classification and Legal Reasoning}: The dataset structure supports the development of models that can predict multiple, interconnected legal violations for a single code fragment.
    \item \emph{Explainable Compliance Detection}: The \textbf{annotation\_note} serves as a ground truth for evaluating the quality of model-generated legal explanations.
\end{enumerate}

By balancing provenance, technical detail, and legal insight, GDPR-Bench-Android establishes a robust and extensible foundation for research at the intersection of software engineering and data protection law.

\subsection{Dataset Statistical Analysis}
\label{sec:dataset_stats}

With the expanded dataset construction and schema established, we present a rigorous quantitative analysis of its \textbf{1951} annotated violations. This analysis investigates the distribution of violations across applications, the prevalence of specific GDPR Articles, file type diversity, and snippet characteristics.

\subsubsection{Application-Level Distribution}
\begin{figure}[tbp]
    \centering
    \begin{subfigure}[b]{1\textwidth}
        \centering
        \includegraphics[width=\textwidth]{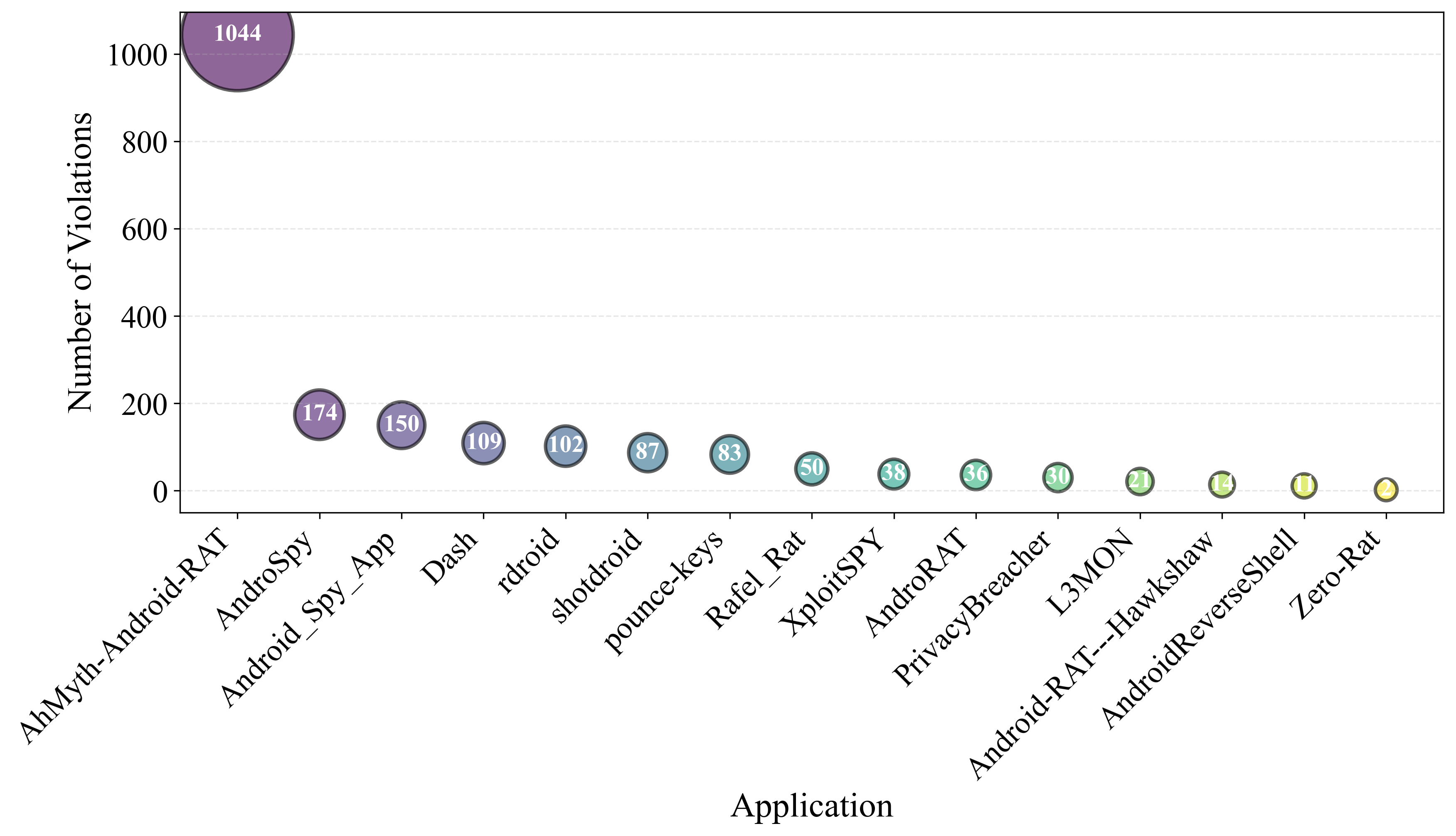}
        \caption{Bubble Chart of Violations per Application (15 Applications)}
        \label{fig:app_bubble_new}
    \end{subfigure}
    \vfill
    \begin{subfigure}[b]{0.8\textwidth}
        \centering
        \includegraphics[width=\textwidth]{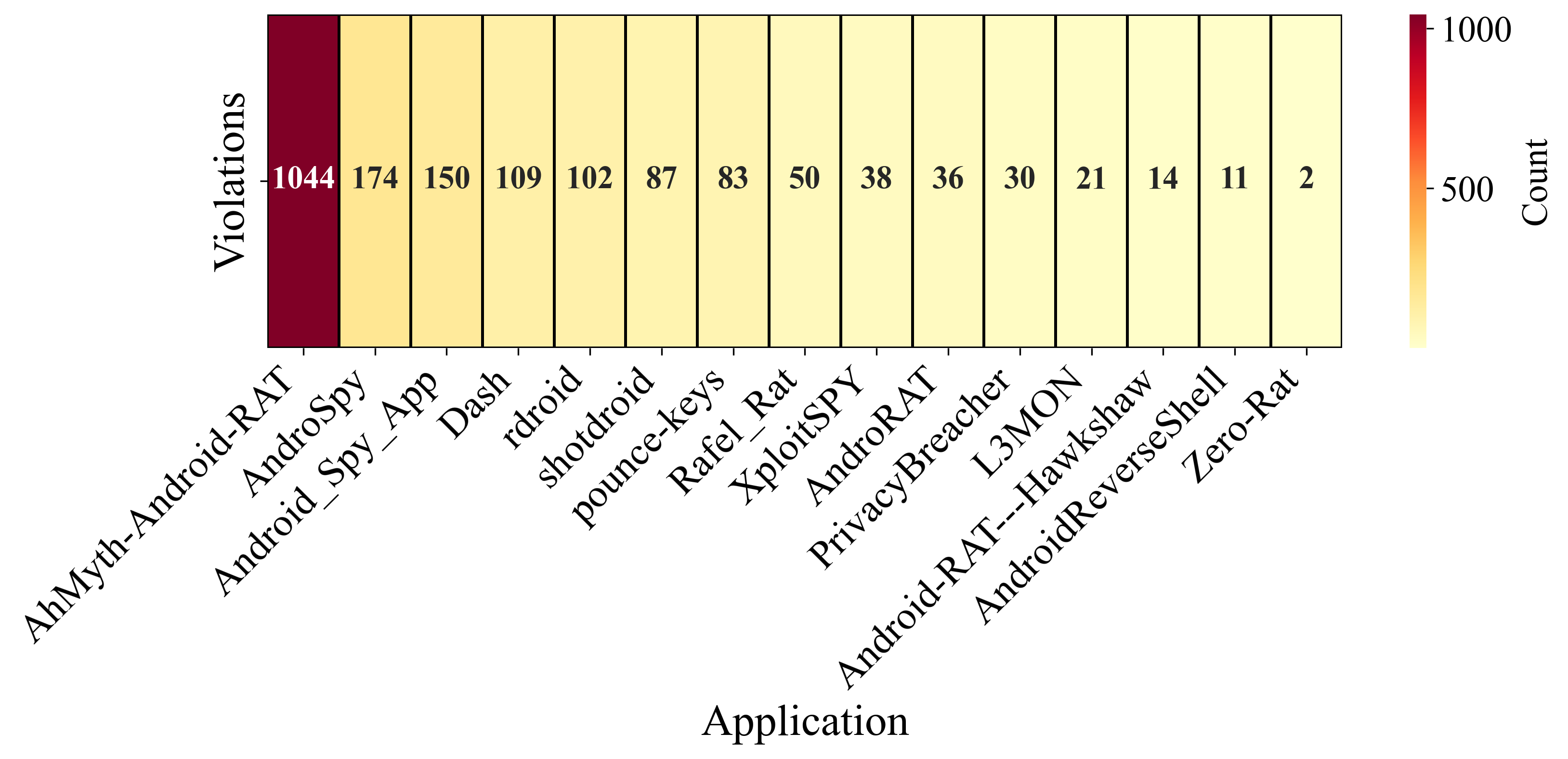}
        \caption{Heatmap of Violations per Application (15 Applications)}
        \label{fig:app_heatmap_new}
    \end{subfigure}
    \caption{GDPR violation distribution across the 15 applications in the expanded dataset. The visualizations show a heavy-tailed distribution, with \textbf{AhMyth-Android-RAT} contributing the majority of instances (1044).}
    \label{fig:applications_comparison}
\end{figure}

Our expanded dataset now includes 15 applications. As shown in Figure~\ref{fig:applications_comparison}, the distribution of violations is heavily skewed. The \textbf{AhMyth-Android-RAT} repository alone accounts for 1044 instances (53.51\%) of all breaches. The next three most prolific applications, \textbf{AndroSpy} (174 instances, 8.92\%), \textbf{Android\_Spy\_App} (150 instances, 7.69\%), and \textbf{Dash} (109 instances, 5.59\%), collectively contribute another 22.2\%.

This distribution is a deliberate result of our corpus selection strategy, addressing reviewer concerns about representativeness. Applications explicitly designed for surveillance (e.g., RATs and spy apps) provide a rich and essential source of unambiguous, non-trivial violations. This is crucial for establishing a robust ground truth for a benchmark.

\subsubsection{Violation Granularity}

Our 1951 annotations are almost evenly split between precise, single-line violations and more complex multi-line violations. As detailed in Table~\ref{tab:granularity_distribution_new}, multi-line snippets account for 994 records (50.95\%), while single-line snippets account for 957 records (49.05\%). This balance ensures the benchmark evaluates both a model's ability to spot localized API misuse (single-line) and its capacity to understand violations embedded in a larger code context (multi-line).

\begin{table}[H]
 \centering
 \caption{Violation granularity distribution (Total: 1951)}
 \label{tab:granularity_distribution_new}
 \begin{tabular}{@{}lrr@{}}
    \toprule
    Granularity & \# Instances & Percentage \\
    \midrule
    Multi-line Level & 994 & 50.95\% \\
    Single-line Level & 957 & 49.05\% \\
    \bottomrule
 \end{tabular}
\end{table}

\subsubsection{Prevalence of GDPR Articles}

\begin{figure}[tbp]
    \centering
    \begin{subfigure}[b]{1\textwidth}
        \centering
        \includegraphics[width=\textwidth]{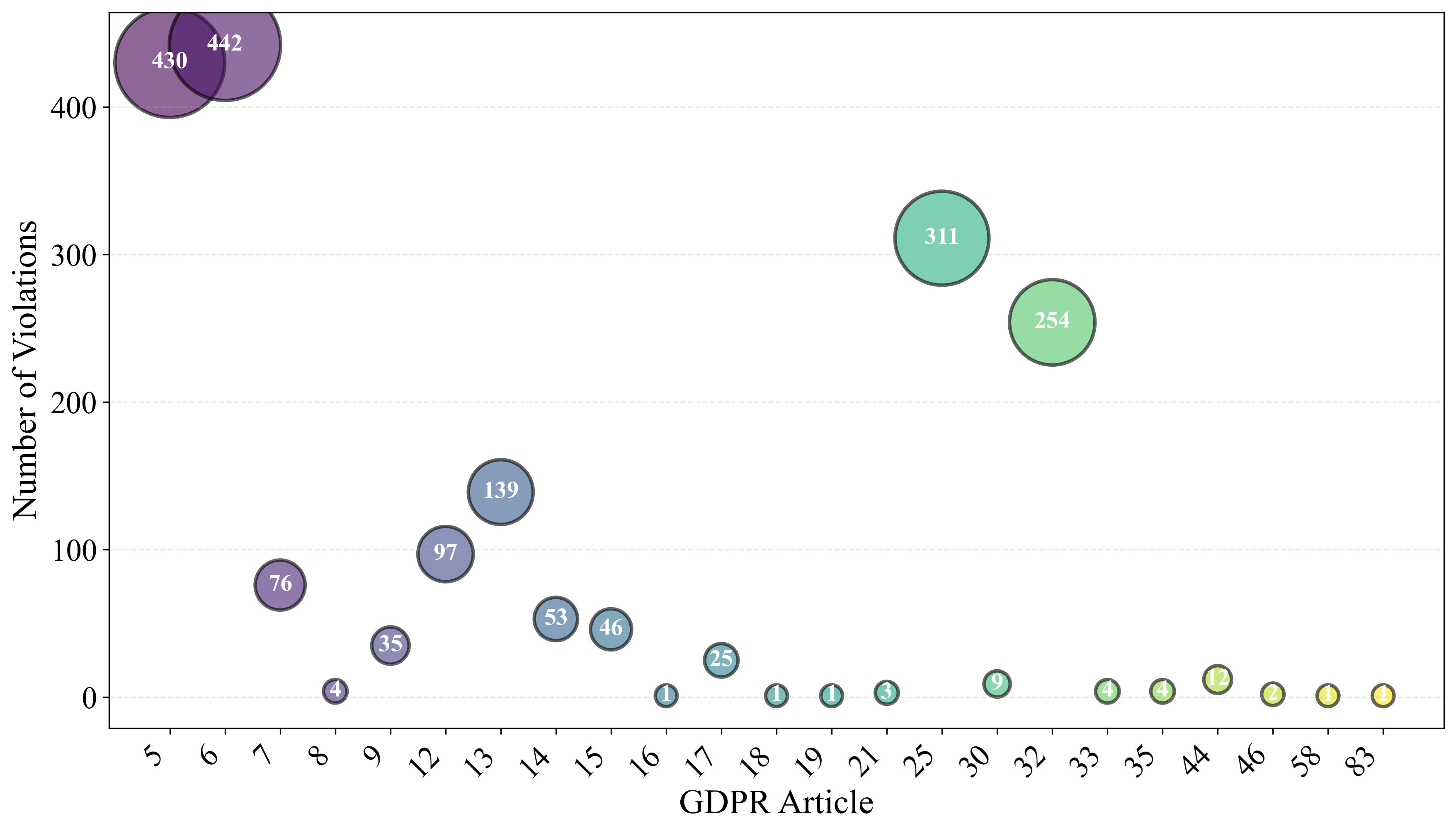}
        \caption{Bubble Chart of Violations per GDPR Article}
        \label{fig:articles_bubble_new}
    \end{subfigure}
    \vfill
    \begin{subfigure}[b]{0.8\textwidth}
        \centering
        \includegraphics[width=\textwidth]{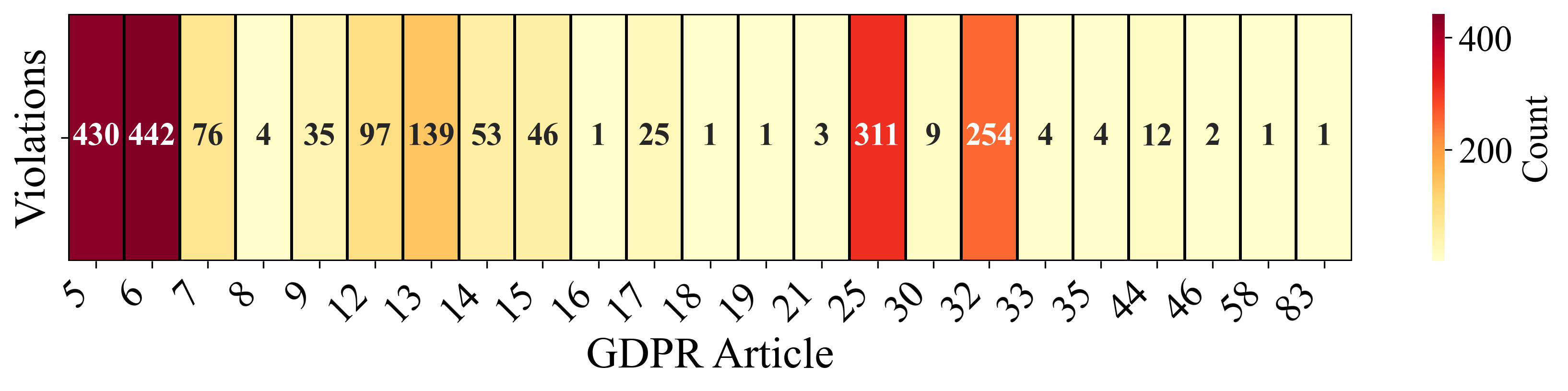}
        \caption{Heatmap of Violations per GDPR Article}
        \label{fig:articles_heatmap_new}
    \end{subfigure}
    \caption{Distribution of violations across all 23 GDPR articles. Articles 6 (442), 5 (430), 25 (311), and 32 (254) are the most frequent. This concentration highlights key compliance failure patterns.}
    \label{fig:articles_comparison}
\end{figure}

The 1951 violations span 23 distinct GDPR Articles. The distribution, shown in Figure~\ref{fig:articles_comparison}, is highly concentrated. The four most frequently breached articles account for over 73\% of all violations:
\begin{itemize}[noitemsep]
    \item \textbf{Article 6} (Lawfulness of processing): 442 instances (22.66\%)
    \item \textbf{Article 5} (Principles of processing): 430 instances (22.04\%)
    \item \textbf{Article 25} (Data protection by design and by default): 311 instances (15.94\%)
    \item \textbf{Article 32} (Security of processing): 254 instances (13.02\%)
\end{itemize}
This concentration highlights that the core challenges in the analyzed codebases relate to fundamental principles: lacking a lawful basis for processing (Art. 6) (e.g., accessing contacts or location without a prior consent check), failing to be transparent (Art. 5) e.g., using a misleading permission rationale to obtain consent, such as describing a keylogger as a "System Settings" service ), and failing to build in security and privacy safeguards from the start (Art. 25, Art. 32) (e.g., transmitting data over unencrypted connections or storing passwords in plain text in SharedPreferences). In contrast, articles concerning specific data subject rights, such as Article 16 (Right to rectification) or Article 18 (Right to restriction), are rare (1 instance each). This real-world skew presents a challenging long-tail distribution problem for automated models.

\subsubsection{File-Type Distribution}
The expanded dataset is highly multilingual. As shown in Table~\ref{tab:file_ext_distribution_new}, violations are not confined to traditional Android (Java/Kotlin) files. JavaScript (\textbf{.js}) files, often part of a RAT's server-side or web-based components, are the most common source of violations (528 instances, 27.06\%). This is followed by configuration files (\textbf{.json}, 24.30\%) and Android's native \textbf{.java} (15.27\%) and \textbf{.kt} (12.51\%) files. The inclusion of \textbf{.cs} (C\#, 8.92\%) and \textbf{.php} (6.46\%) files further underscores the diverse, multi-stack nature of modern privacy violations.

\begin{table}[t]
 \centering
 \caption{Code snippet file-type distribution (Total: 1951)}
 \label{tab:file_ext_distribution_new}
 \begin{tabular}{@{}lrr@{}}
    \toprule
    Extension & \# Instances & Percentage \\
    \midrule
    \textbf{.js} & 528 & 27.06\% \\
    \textbf{.json} & 474 & 24.30\% \\
    \textbf{.java} & 298 & 15.27\% \\
    \textbf{.kt} & 244 & 12.51\% \\
    \textbf{.cs} & 174 & 8.92\% \\
    \textbf{.php} & 126 & 6.46\% \\
    \textbf{.xml} & 63 & 3.23\% \\
    \textbf{.html} & 26 & 1.33\% \\
    \textbf{.py} & 17 & 0.87\% \\
    \textbf{.h} & 1 & 0.05\% \\
    \bottomrule
 \end{tabular}
\end{table}

\subsubsection{Snippet and Annotation Lengths}
Finally, Table~\ref{tab:length_stats_full} summarizes the length characteristics of the code snippets and their corresponding legal explanations. Code snippets have a mean length of 171.01 characters, while the expert-written annotation notes are longer, with a mean of 224.38 characters. This indicates that the legal rationale for a violation is often more complex than the code that implements it. Models targeting GDPR-Bench-Android must be able to handle this variability.

\begin{table}[t]
 \centering
 \caption{Snippet and annotation length statistics (characters) (N=1951)}
 \label{tab:length_stats_full}
 \begin{tabular}{@{}lrrrrr@{}}
    \toprule
     & Min & Max & Mean & Median & Std. Dev. \\
    \midrule
    Snippet length & 12 & 1717 & 171.01 & 95.0 & 206.58 \\
    Annotation length & 60 & 684 & 224.38 & 208.0 & 90.00 \\
    \bottomrule
 \end{tabular}
\end{table}

The insights from this new analysis—the intentional inclusion of high-violation surveillance apps, the balanced mix of single- and multi-line snippets, the high concentration in Articles 5, 6, 25, and 32, and the highly multilingual nature of the violations—confirm that GDPR-Bench-Android provides a challenging and realistic benchmark for evaluating modern automated compliance tools.

\section{Tasks}
\label{sec:tasks}

\subsection{Task Design Rationale}
\label{sec:task_design_rationale}

Real-world GDPR violations manifest at multiple layers of a codebase: missing permission checks can affect entire service classes, improper use of sensitive APIs may compromise modules, and an unguarded API call constitutes a precise line-level breach. To faithfully benchmark the compliance capabilities of \emph{automated methods}, we therefore decompose our evaluation into two complementary tasks:

\begin{itemize}[noitemsep]
  \item \textbf{Task 1: Multi-Granularity Violation Localization.}
    Methods must determine, for each repository commit, which GDPR Articles are violated at \emph{file}-, \emph{module}-, and \emph{line}-level scopes. This task probes hierarchical reasoning over both broad architectural contexts and fine-grained statements.
  \item \textbf{Task 2: Snippet-Level Multi-Label Classification.}
    Methods are presented with isolated code snippets and must assign all applicable Article labels. This task isolates local context comprehension and multi-label prediction.
\end{itemize}

By separating structural localization (Task 1) from standalone snippet classification (Task 2), we can assess the distinct capabilities of different automated paradigms. Task 1 challenges the holistic, long-context reasoning required by agentic approaches, while Task 2 tests the localized pattern-matching strengths vital for both LLMs and symbolic (rule-based) methods. Prior studies in LLM-driven program analysis~\cite{jimenez2024swebenchlanguagemodelsresolve, bigcodebench2024} and automated compliance reasoning suggest that breaking down hierarchical and local tasks improves model interpretability and performance.

\subsection{Dataset Generation Pipeline}
\label{sec:dataset_generation_pipeline}

\begin{figure}[tbp]
    \centering
    \includegraphics[width=\textwidth]{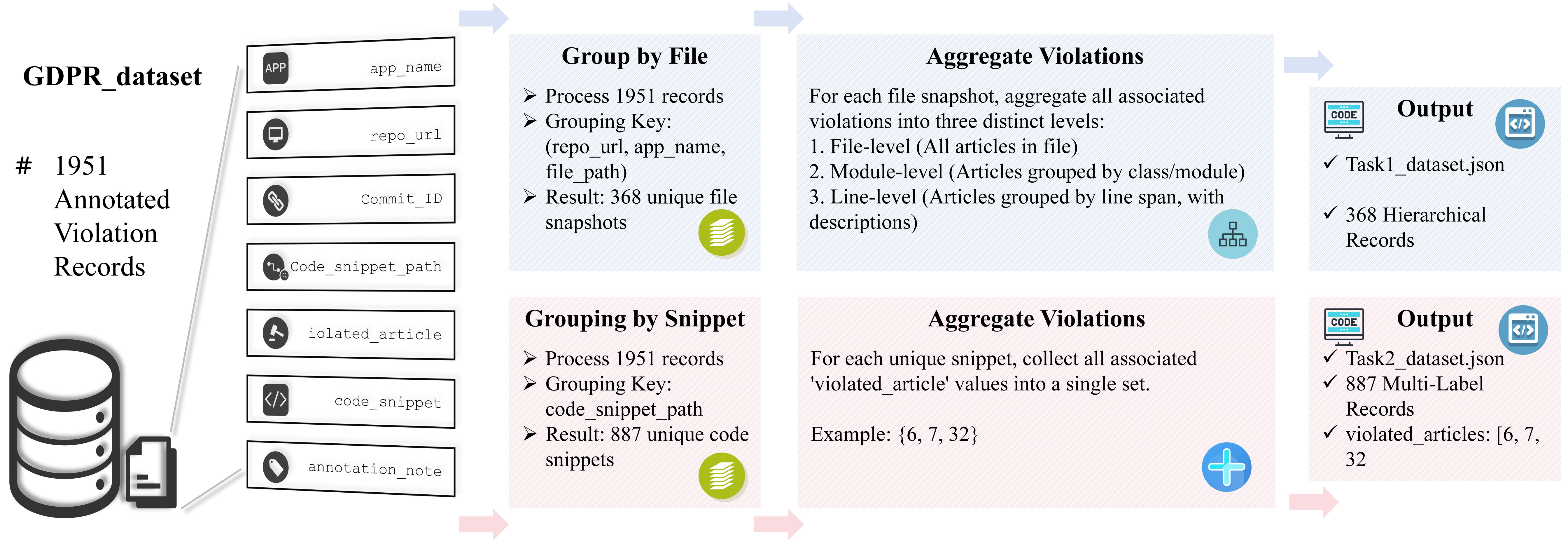}
    \caption{The unified dataset generation pipeline. Both task-specific datasets originate from the same 1951-record source corpus. Task 1 (top branch) groups records by file, creating 368 hierarchical entries. Task 2 (bottom branch) groups records by code snippet, creating 887 multi-label entries.}
    \label{fig:unified_pipeline}
\end{figure}

The datasets for both Task 1 and Task 2 are systematically derived from our single source corpus, which contains \textbf{1951 annotated violation records}. Figure~\ref{fig:unified_pipeline} illustrates the complete data transformation pipeline, showing how this source corpus is processed in parallel to generate the two distinct task datasets. The specific methodologies for each branch (Task 1 and Task 2) are detailed in the following sections.

\subsection{Task 1 Dataset Construction}
\label{sec:task1_dataset}

Task 1's dataset is systematically derived from our comprehensive \textbf{1951-entry} GDPR violation corpus through an automated transformation pipeline that follows established methodologies from SWE-bench~\cite{jimenez2024swebenchlanguagemodelsresolve} and BigCodeBench~\cite{bigcodebench2024}. These frameworks structure datasets around authentic real-world code contexts and hierarchical violation scopes, enabling sophisticated evaluation of model capabilities across multiple levels of code abstraction.


As illustrated in the top branch of Figure~\ref{fig:unified_pipeline}, our systematic approach ensures that each dataset entry corresponds to a single file's comprehensive multi-granularity violation profile. The pipeline transforms the original violation corpus into \textbf{368} file-centric entries that capture regulatory breaches at file, module, and line levels simultaneously.

\paragraph{Load and Group by File}
The transformation process begins by loading the original JSON corpus and systematically grouping violations by the unique tuple \textbf{(repo\_url, app\_name, file\_path)}. This ensures that each distinct file across all repositories forms an independent dataset entry, as depicted in the initial grouping stage of Figure~\ref{fig:unified_pipeline}. This approach yields a set of distinct file-based samples, with each sample representing a complete snapshot of GDPR compliance issues within a specific source file.


\paragraph{Multi-Granularity Violation Aggregation}
Following the extraction phase shown in the Task 1 branch of Figure~\ref{fig:unified_pipeline}, our systematic aggregation methodology constructs three distinct violation collections for each file group, capturing GDPR compliance issues across different levels of code abstraction. This hierarchical approach enables comprehensive evaluation of model performance across various scopes of regulatory analysis:

\begin{itemize}[noitemsep]
  \item \textbf{File-level Violations:}
    Systematically aggregate all Article violations identified within the entire source file, providing a holistic view of the file's compliance status.
  \item \textbf{Module-level Violations:}
    Intelligently infer module or class names from file paths and annotation metadata, systematically grouping Article violations per identified module within the file.
  \item \textbf{Line-level Violations:}
    Precisely parse \textbf{code\_snippet\_path} annotations to extract exact line spans, supporting both single-line ("line X") and multi-line ("lines X–Y") violation specifications. Each line-level entry preserves the expert-authored violation description.
\end{itemize}

This multi-granularity aggregation strategy ensures that models are evaluated across the full spectrum of compliance reasoning, from high-level architectural assessment to precise statement-level violation identification.

\begin{figure}[tbp]
 \centering
 \includegraphics[width=0.8\textwidth]{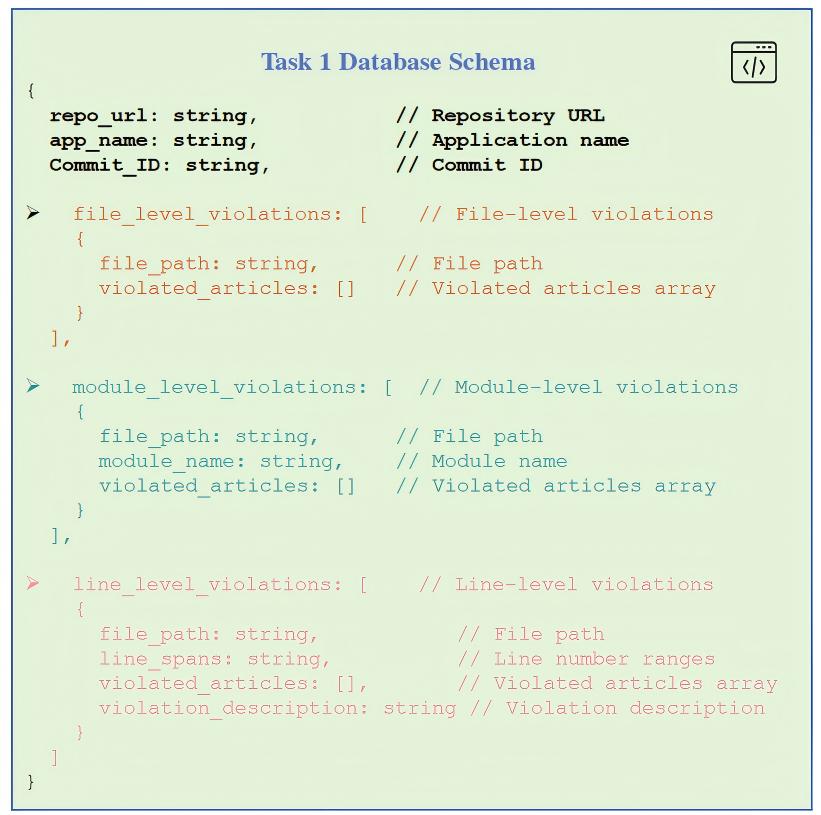}
 \caption{The hierarchical JSON schema for \textbf{task1\_dataset.json}. Each entry contains provenance metadata and three distinct arrays for violations at the file, module, and line levels.}
 \label{fig:task1_schema}
\end{figure}

\paragraph{Structured JSON Generation}
The final transformation stage, as depicted in the output phase of Figure~\ref{fig:unified_pipeline}, serializes each file's violation profile into a structured JSON object. This approach ensures that each dataset entry represents exactly one file's complete multi-granularity GDPR compliance assessment. The comprehensive schema, detailed in Figure~\ref{fig:task1_schema}, demonstrates how each file-centric entry systematically organizes violation information.


\paragraph{Dataset Characteristics and Coverage}
The resulting Task 1 dataset contains \textbf{368} file-centric records and exhibits comprehensive coverage across all violation granularities. Each entry corresponds to a unique file within a specific commit from one of the \textbf{15} applications. As detailed in the schema (Figure~\ref{fig:task1_schema}), each violation category maintains specific structural properties that support different types of compliance analysis.

Statistical analysis of the underlying 1951 violations reveals predominant breaches in fundamental GDPR principles, particularly Articles 5 (processing principles), 6 (lawful basis for processing), 25 (data protection by design), and 32 (security of processing). The fine-grained line-level annotations demonstrate substantial depth, with violation descriptions averaging 224.38 characters in length. This rich contextual information provides essential semantic detail for training and evaluating models on complex regulatory reasoning tasks.

The multi-granularity structure, as exemplified in Figure~\ref{fig:task1_schema}, enables researchers to assess model performance across different scopes of compliance analysis, from broad file-level regulatory assessment to precise line-level violation localization and explanation. This systematic transformation approach ensures that Task 1 provides a rigorous foundation for evaluating automated methods in multi-scale GDPR compliance detection.

\subsection{Task 2 Dataset Construction}

To complement the hierarchical localization capabilities evaluated in Task 1, Task 2 provides a streamlined, snippet-level multi-label classification benchmark. This task design draws inspiration from privacy labeling frameworks such as CLAUDETTE~\cite{lippi2019claudette}, where granular multi-label tasks reveal nuanced compliance gaps.


As illustrated in the bottom branch of Figure~\ref{fig:unified_pipeline}, the Task 2 dataset construction process transforms the original \textbf{1951-record} corpus into a flat, snippet-centric format optimized for multi-label classification, resulting in \textbf{887} unique snippet records.

\paragraph{Manual Annotation and Multi-Label Aggregation}
We did not perform additional annotation for Task 2. Instead, we used the initial labeling from our 1951-record dataset. As explained in Section~\ref{sec:dataset_explanation} and Listing~\ref{lst:example_json}, our annotation process often generated multiple records for a single code snippet, with each record citing a different \textbf{violated\_article}.

To create the Task 2 dataset, our aggregation mechanism groups all 1951 records by their \textbf{code\_snippet\_path} identifier. This ensures that each unique code location becomes a single entry in the Task 2 dataset. The \textbf{violated\_articles} from all corresponding records are then aggregated into a single set. This approach naturally handles cases where individual code snippets violate multiple GDPR provisions simultaneously, creating a realistic multi-label classification scenario.

\begin{figure}[tbp]
 \centering
 \includegraphics[width=\textwidth]{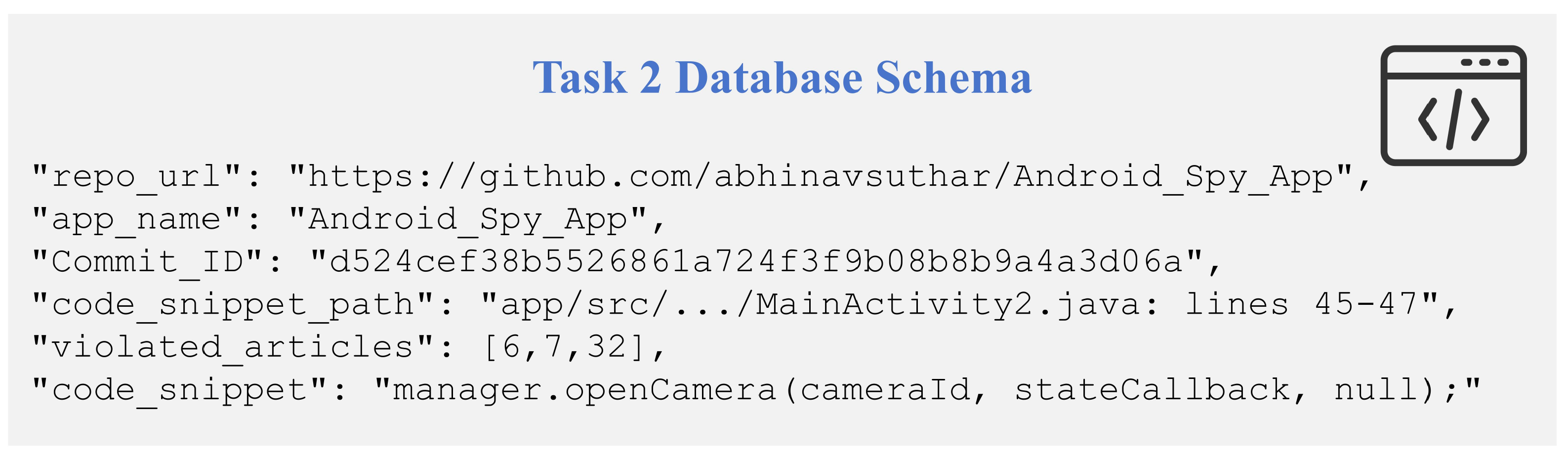}
 \caption{An example record from the \textbf{task2\_dataset.json} schema. Note that \textbf{violated\_articles} is an array of integers, representing the multi-label ground truth for the given \textbf{code\_snippet}.}
 \label{fig:task2_schema}
\end{figure}

\paragraph{Dataset Generation and Schema Design}
The final stage transforms the grouped annotations into a structured JSON format. As shown in Figure~\ref{fig:task2_schema}, each record maintains essential provenance information while focusing on the core classification task. The \textbf{violated\_articles} field is now an array of integers representing the complete set of violations for that snippet.


\paragraph{Technical Implementation Details}
The transformation from the original 1951-record corpus to the Task 2 format involves a straightforward grouping logic. We initialize an empty dictionary to hold the grouped data. We then iterate through each of the 1951 entries in the original dataset. For each entry, we use its \textbf{code\_snippet\_path} as a unique key. If this key has not been seen before, we create a new record for it and copy over the provenance fields (\textbf{repo\_url}, \textbf{app\_name}, \textbf{Commit\_ID}, \textbf{code\_snippet}). We then add the entry's \textbf{violated\_article} (a single integer) to a \textbf{set} associated with this key. If the key *has* been seen before, we simply add the new \textbf{violated\_article} to the existing set. This process automatically consolidates all violations for a single snippet and removes duplicate article numbers, resulting in the final multi-label dataset.

\paragraph{Output Generation and Quality Assurance}
Each processed group is systematically transformed into a standardized JSON object. The \textbf{violated\_articles} field, which was a single integer in the original 1951 records, is now a sorted array of all applicable GDPR provisions for that snippet. The complete dataset is serialized to \textbf{task2\_dataset.json}, creating a robust foundation for multi-label classification evaluation.

\subsection{Summary of Task Construction}
By systematically deriving two complementary datasets from the same foundational violation corpus—one hierarchical (Task 1) and one flat (Task 2)—we establish a comprehensive benchmark that rigorously exercises both global architectural reasoning and local compliance detection capabilities in GDPR violation identification. Task 1 challenges models to reconcile broad contextual understanding with precise statement-level breach detection, while Task 2 evaluates fine-grained, multi-label classification performance on isolated code fragments. This dual-task framework provides a robust foundation for rigorous evaluation and systematic comparison of diverse \textbf{automated methods} in the critical domain of automated legal compliance assessment for software engineering.

\section{Experimental Evaluation}
\label{sec:experiments}

In this section, we describe the experimental methodology used to assess the ability of different automated methods to detect GDPR violations in source code. We begin by detailing the evaluation settings, including the choice of models and baseline methods, hardware/software environment, and inference configuration; we then present the metrics used to quantify performance at multiple granularity levels.

\subsection{Evaluation Settings}
\label{subsec:eval-settings}

All experiments for baseline LLMs were executed under a unified, zero-shot protocol~\cite{brown2020language, schick2021exploiting, sanh2022multitask} to ensure fair comparison. Specialized methods (RAG, ReAct) and the formal method use their own distinct protocols as described below. Below we specify the models under evaluation, the computing platform, the prompt and inference parameters, and the formal definitions of our evaluation metrics.

\subsubsection{Models and Methods under Evaluation}
\label{subsubsec:models-and-methods}

We evaluate a total of 11 distinct approaches, comprising eight baseline LLMs and three additional methods to provide a comprehensive comparison against symbolic, retrieval-augmented, and agentic paradigms.

\paragraph{Baseline Large Language Models}
We benchmark eight representative LLMs spanning both open- and closed-source offerings:
\begin{itemize}
    \item \textbf{Open-source}: Deepseek-r1, Qwen2.5-72B, Gemini-2.5-pro-preview-06-05-thinking, Gemini-2.5-pro-thinking
    \item \textbf{Closed-source}: Claude-3.7-Sonnet-20250219, Claude-Sonnet-4.5-20250929, GPT-4o, O1
\end{itemize}
Each of these baseline models is invoked via its official RESTful API using the unified zero-shot prompt template (see Section \ref{subsubsec:prompting}) in a zero-shot fashion.

\paragraph{Specialized and Formal Methods}
To establish a comprehensive evaluation landscape, we implement and evaluate three additional methods representing distinct paradigms in automated code analysis.

\begin{itemize}
    \item \textbf{Formal-AST (Formal Method):}
    To contrast the performance of LLMs against traditional static analysis, we implemented a formal, rule-based method inspired by established symbolic techniques~\cite{arzt2014flowdroid, li2015iccta}. This method does not use any LLM. It first uses a multi-language parser to build an AST for the given code. It then traverses the AST to extract facts and populates a set of formal predicates (e.g., \textbf{CollectsDataPredicate(DEVICE\_ID\_DATA)} is \textbf{True}). Finally, it evaluates over 35 formal logic rules (e.g., \textbf{CollectsDataPredicate AND NOT HasConsentPredicate} $\rightarrow$ \textbf{Violation(Article 6)}) derived from our dataset and GDPR text to determine violations.

    \item \textbf{RAG (Retrieval-Augmented Method):}
    The RAG method enhances a base LLM with external knowledge. During initialization, it constructs a vector knowledge base from two sources: (1) textual definitions of all GDPR articles, and (2) all 1,900+ annotated code-violation pairs from our dataset. When analyzing a new code snippet, it first retrieves the most relevant documents (e.g., article definitions and semantically similar code examples from the dataset) and injects this context into the LLM's prompt, enabling in-context learning~\cite{brown2020language}.

    \item \textbf{ReAct (Agentic Method):}
    To investigate the capabilities of more recent agentic architectures in complex reasoning tasks, we implemented a ReAct (Reasoning and Acting) agent~\cite{yao2023react}. This agent uses an LLM (GPT-4o) as a reasoning "brain" that can iteratively use a set of custom tools. These tools allow the agent to: (1) \textbf{gdpr\_lookup(article\_id)} to read article definitions, (2) \textbf{code\_search(keyword)} to find sensitive APIs, and (3) \textbf{rule\_check(code\_snippet)} to run a simple static analysis. The agent reasons about the code, decides which tool to use, observes the output, and repeats this loop until it reaches a final conclusion.
\end{itemize}

\subsubsection{Hardware and Software Environment}
Our evaluation was carried out on a dedicated workstation with the following configuration:
\begin{itemize}
  \item \textbf{Operating System}: Microsoft Windows 11 Professional (build 10.0.22631)
  \item \textbf{Processor}: Intel Core i9-12900H (14 cores, 20 threads @ 2.50 GHz)
  \item \textbf{Memory}: 16 GB DDR4 RAM
  \item \textbf{BIOS}: American Megatrends UEFI G733ZW.324 (2023-02-21)
  \item \textbf{Software}: Python 3.10.12 with {requests}, {numpy}, {scikit-learn}, {matplotlib}
\end{itemize}

\subsubsection{Prompting and Inference Configuration}
\label{subsubsec:prompting}
We employ a fixed zero-shot prompt for all baseline LLM experiments. The prompt template is shown in Figure~\ref{fig:prompt}.

\begin{figure}[tbp]
\begin{verbatim}
You are a GDPR compliance expert. Your task is to determine which GDPR 
articles are violated by the following code snippet.

GDPR Article Meanings:
- Article 5: Principles of processing
- Article 6: Lawfulness of processing
- Article 7: Conditions for consent
... (definitions for all relevant articles) ...

Instructions:
- Carefully analyze the code snippet.
- Only output the violated GDPR article numbers, separated by commas (e.g.,5,6,32).
- If there is no violation, output exactly 0.

Code snippet:
<Code Snippet Here>
\end{verbatim}
\caption{Zero-shot prompt template for baseline LLM evaluation.}
\label{fig:prompt}
\end{figure}

The specialized methods (RAG, ReAct) and the Formal-AST method use their own inference procedures and prompts as detailed in Section~\ref{subsubsec:models-and-methods}.

Inference parameters were held constant across all LLM-based methods:
\begin{description}
  \item[Temperature:] 0.0  
  \item[Top-$p$:] 1.0  
  \item[Maximum response length:] 512 tokens  
  \item[Number of completions:] 1  
\end{description}

\subsubsection{Evaluation Metrics}

We employ task-specific evaluation metrics with rigorous mathematical definitions to assess both hierarchical violation localization and multi-label classification performance.

\paragraph{Task 1: Multi-granularity Violation Localization}
For ranking-based evaluation, we measure the method's ability to correctly rank GDPR articles across different code granularities. For each example $i$ at granularity $g \in \{\text{file, module, line}\}$, let $r_i^{(g)}$ be the rank of the first correct GDPR article in the method's predictions. The Accuracy@$k$ metric is defined as:

\[
\mathrm{Accuracy}@k_g = \frac{1}{N}\sum_{i=1}^{N}\mathbf{1}\{r_i^{(g)} \le k\}, \quad k = 1,2,\ldots,5
\]

where $N$ is the number of test instances and $\mathbf{1}\{\cdot\}$ is the indicator function. This metric captures whether the correct violation appears within the top-$k$ predictions, enabling evaluation of both precision (k=1) and recall depth (k=5).

\paragraph{Task 2: Snippet-level Multi-label Classification}
For multi-label evaluation, we employ standard classification metrics adapted for the regulatory compliance domain. Let $y_{i,c} \in \{0,1\}$ denote the ground-truth label for article $c$ in snippet $i$, and $\hat{y}_{i,c} \in \{0,1\}$ the method's prediction. We compute four key metrics:

\[
\mathrm{Precision}_{\mathrm{macro}} = \frac{1}{C}\sum_{c=1}^{C}\mathrm{Precision}_c, \quad
\mathrm{Recall}_{\mathrm{macro}} = \frac{1}{C}\sum_{c=1}^{C}\mathrm{Recall}_c, \quad
\mathrm{F1}_{\mathrm{macro}} = \frac{1}{C}\sum_{c=1}^{C}\mathrm{F1}_c
\]

\[
\mathrm{Accuracy} = \frac{1}{N \cdot C}\sum_{i=1}^{N}\sum_{c=1}^{C}\mathbf{1}\{\hat{y}_{i,c} = y_{i,c}\}
\]

where $C$ represents the total number of GDPR articles. Macro-averaging ensures balanced evaluation across all articles, preventing frequent violations from overshadowing rarer but equally critical compliance issues. All metrics are implemented in our evaluation framework (\textbf{evaluate\_model.py}).

\subsection{Experimental Results}
\label{subsec:results}
We evaluate the performance of all 11 methods on both multi-granularity violation localization (Task 1) and snippet-level multi-label classification (Task 2). This evaluation now includes eight baseline LLMs, our Formal-AST method, a retrieval-augmented method (RAG), and an agentic method (ReAct).

\subsubsection{Task 1: Multi-granularity Violation Localization}
\label{subsubsec:task1}

We evaluate each method's performance on hierarchical GDPR violation localization across file, module, and line granularities using Accuracy@$k$ ($k=1 \dots 5$). The following analysis compares the performance of baseline LLMs against the Formal-AST, RAG, and ReAct methods at each scope, with results presented in Tables~\ref{tab:task1_file_results}, \ref{tab:task1_module_results}, and \ref{tab:task1_line_results}.

\paragraph{File-level Violation Localization}

At the file-level, which requires understanding the broadest context, the agentic ReAct method demonstrates superior performance. As shown in Table~\ref{tab:task1_file_results}, ReAct achieves the highest Accuracy@1 (17.38\%), substantially outperforming all baseline LLMs. This result supports the hypothesis that agentic approaches, which can iteratively use tools to analyze long contexts, are better suited for holistic code analysis. The best performing baseline LLM is Deepseek-r1 with an Accuracy@1 of 11.55\%. The RAG method (7.34\% @1) shows a notable improvement over its base model, GPT-4o (3.86\% @1), suggesting that retrieval is beneficial even in large contexts. The Formal-AST method (9.79\% @1) performs competitively with several LLMs, indicating that its rule-based patterns are effective even at a file-wide scope, but it shows limited improvement at deeper recall levels (18.36\% @5).

\begin{table}[t]
\centering
\caption{Task 1: File-level Accuracy@$k$ performance across all methods.}
\label{tab:task1_file_results}
\small
\begin{tabular}{lccccc}
\toprule
\textbf{Method} & \textbf{@1} & \textbf{@2} & \textbf{@3} & \textbf{@4} & \textbf{@5} \\
\midrule
\multicolumn{6}{l}{\textit{Baseline LLMs}} \\
claude-3-7-sonnet-20250219 & 0.0567 & 0.1355 & 0.1690 & 0.1998 & 0.2369 \\
claude-sonnet-4-5-20250929 & 0.0621 & 0.1486 & 0.1798 & 0.2146 & 0.2465 \\
deepseek-r1 & 0.1155 & 0.1540 & 0.1933 & 0.2365 & 0.2621 \\
gemini-2.5-pro-preview-06-05-thinking & 0.0954 & 0.1934 & 0.2244 & 0.2436 & 0.2836 \\
gemini-2.5-pro-thinking & 0.0915 & 0.1855 & 0.2164 & 0.2359 & 0.2736 \\
gpt-4o & 0.0386 & 0.0938 & 0.1209 & 0.1391 & 0.1571 \\
o1 & \underline{0.0262} & 0.0759 & 0.0969 & 0.1153 & 0.1381 \\
qwen2.5-72b-instruct & 0.0274 & \underline{0.0701} & \underline{0.0954} & \underline{0.1116} & \underline{0.1168} \\
\midrule
\multicolumn{6}{l}{\textit{Specialized and Formal Methods}} \\
Formal-AST & 0.0979 & 0.1263 & 0.1573 & 0.1779 & 0.1836 \\
RAG & 0.0734 & 0.1062 & 0.1378 & 0.1514 & 0.1604 \\
ReAct & \textbf{0.1738} & \textbf{0.2138} & \textbf{0.2788} & \textbf{0.3936} & \textbf{0.3958} \\
\bottomrule
\end{tabular}
\end{table}

\paragraph{Module-level Violation Localization}

At the module level, the performance trends are similar to the file-level, as shown in Table~\ref{tab:task1_module_results}. The ReAct method continues to lead with the highest Accuracy@1 (17.13\%), again demonstrating its effectiveness in handling large but somewhat focused code contexts. Among baseline LLMs, `claude-sonnet-4-5-20250929` (9.60\% @1) and `deepseek-r1` (9.38\% @1) are the top performers. The Formal-AST method (9.73\% @1) remains robust, outperforming most LLMs in top-1 accuracy, which suggests its rules are well-suited for logical code units like classes. The RAG method (8.30\% @1) again offers a significant improvement over its base model, reinforcing the observation that retrieval is beneficial for contextual analysis.

\begin{table}[t]
\centering
\caption{Task 1: Module-level Accuracy@$k$ performance across all methods.}
\label{tab:task1_module_results}
\small
\begin{tabular}{lccccc}
\toprule
\textbf{Method} & \textbf{@1} & \textbf{@2} & \textbf{@3} & \textbf{@4} & \textbf{@5} \\
\midrule
\multicolumn{6}{l}{\textit{Baseline LLMs}} \\
claude-3-7-sonnet-20250219 & 0.0607 & 0.1422 & 0.1763 & 0.2147 & 0.2561 \\
claude-sonnet-4-5-20250929 & 0.0960 & \textbf{0.2054} & 0.2450 & 0.2958 & 0.3295 \\
deepseek-r1 & 0.0938 & 0.1309 & 0.1632 & 0.1927 & 0.2121 \\
gemini-2.5-pro-preview-06-05-thinking & 0.0800 & 0.1633 & 0.1935 & 0.2143 & 0.2553 \\
gemini-2.5-pro-thinking & 0.0686 & 0.1512 & 0.1830 & 0.2087 & 0.2475 \\
gpt-4o & 0.0453 & 0.1077 & 0.1370 & 0.1704 & 0.1841 \\
o1 & 0.0355 & 0.0976 & 0.1313 & 0.1569 & 0.1783 \\
qwen2.5-72b-instruct & \underline{0.0301} & \underline{0.0823} & \underline{0.1095} & \underline{0.1266} & \underline{0.1368} \\
\midrule
\multicolumn{6}{l}{\textit{Specialized and Formal Methods}} \\
Formal-AST & 0.0973 & 0.1256 & 0.1564 & 0.1769 & 0.1826 \\
RAG & 0.0830 & 0.1115 & 0.1445 & 0.1581 & 0.1605 \\
ReAct & \textbf{0.1713} & \textbf{0.2083} & \textbf{0.2732} & \textbf{0.3893} & \textbf{0.3936} \\
\bottomrule
\end{tabular}
\end{table}

\paragraph{Line-level Violation Localization}

At the fine-grained line level, where context is minimal, the performance landscape shifts dramatically (Table~\ref{tab:task1_line_results}). Baseline LLMs excel, with `qwen2.5-72b-instruct` achieving an impressive 61.60\% Accuracy@1. This suggests that for short, localized snippets, LLMs' pre-trained knowledge is highly effective at recognizing explicit violation patterns. The RAG method (56.87\% @1) performs on par with its base model, GPT-4o (57.45\% @1), suggesting that for highly localized snippets, the retrieved context might not add significant value over the model's inherent knowledge. The ReAct agent (50.84\% @1) is competitive but does not lead, likely because its iterative tool-use is less advantageous when the entire context is already small. The Formal-AST method's performance is notably poor (1.86\% @1), revealing its main weakness: its rules are designed for broader contexts (e.g., missing permission checks) and struggle to infer violations from isolated lines.

\begin{table}[t]
\centering
\caption{Task 1: Line-level Accuracy@$k$ performance across all methods.}
\label{tab:task1_line_results}
\small
\begin{tabular}{lccccc}
\toprule
\textbf{Method} & \textbf{@1} & \textbf{@2} & \textbf{@3} & \textbf{@4} & \textbf{@5} \\
\midrule
\multicolumn{6}{l}{\textit{Baseline LLMs}} \\
claude-3-7-sonnet-20250219 & 0.3888 & 0.5601 & 0.6242 & 0.6634 & 0.6811 \\
claude-sonnet-4-5-20250929 & 0.4001 & 0.6598 & 0.7272 & 0.7580 & 0.7757 \\
deepseek-r1 & \underline{0.2744} & \underline{0.3664} & \underline{0.3971} & \underline{0.4155} & \underline{0.4188} \\
gemini-2.5-pro-preview-06-05-thinking & 0.3615 & 0.5690 & 0.6543 & 0.6922 & 0.7390 \\
gemini-2.5-pro-thinking & 0.3531 & 0.5717 & 0.6654 & 0.7095 & 0.7469 \\
gpt-4o & 0.5745 & 0.6448 & 0.6528 & 0.6597 & 0.6625 \\
o1 & 0.4760 & 0.6457 & 0.6641 & 0.6674 & 0.6689 \\
qwen2.5-72b-instruct & \textbf{0.6160} & \textbf{0.7489} & \textbf{0.7838} & \textbf{0.7973} & \textbf{0.8003} \\
\midrule
\multicolumn{6}{l}{\textit{Specialized and Formal Methods}} \\
Formal-AST & 0.0186 & 0.0298 & 0.0416 & 0.0562 & 0.0607 \\
RAG & 0.5687 & 0.6230 & 0.6402 & 0.6442 & 0.6449 \\
ReAct & 0.5084 & 0.5893 & 0.6313 & 0.6910 & 0.7012 \\
\bottomrule
\end{tabular}
\end{table}

\subsubsection{Task 2: Snippet-level Multi-label Classification}
\label{subsubsec:task2-detailed}
Task 2 evaluates the more challenging task of identifying the complete set of GDPR articles violated within a single code snippet. As shown in Table~\ref{tab:task2_results_all}, all methods find this multi-label classification significantly more difficult than ranking-based localization.

\begin{table}[t]
\centering
\caption{Task 2: Snippet-level multi-label classification performance across all methods.}
\label{tab:task2_results_all}
\small
\begin{tabular}{lcccc}
\toprule
\textbf{Method} & \textbf{Accuracy} & \textbf{Macro-Precision} & \textbf{Macro-Recall} & \textbf{Macro-F1} \\
\midrule
\multicolumn{5}{l}{\textit{Baseline LLMs}} \\
claude-3-7-sonnet-20250219 & 0.0486 & 0.0394 & 0.0200 & 0.0234 \\
claude-sonnet-4-5-20250929 & \textbf{0.1313} & 0.0610 & \textbf{0.0653} & \textbf{0.0575} \\
deepseek-r1 & 0.0608 & 0.0271 & 0.0269 & 0.0259 \\
gemini-2.5-pro-preview-06-05-thinking & 0.0805 & 0.0428 & 0.0296 & 0.0328 \\
gemini-2.5-pro-thinking & 0.0808 & 0.0527 & 0.0319 & 0.0361 \\
gpt-4o & 0.0783 & 0.0559 & 0.0245 & 0.0318 \\
o1 & 0.0485 & 0.0525 & 0.0214 & 0.0248 \\
qwen2.5-72b-instruct & \underline{0.0229} & 0.0551 & 0.0089 & 0.0129 \\
\midrule
\multicolumn{5}{l}{\textit{Specialized and Formal Methods}} \\
Formal-AST & 0.0233 & \underline{0.0239} & \underline{0.0071} & \underline{0.0095} \\
RAG & 0.0694 & \textbf{0.0710} & 0.0330 & 0.0416 \\
ReAct & 0.0650 & 0.0247 & 0.0344 & 0.0161 \\
\bottomrule
\end{tabular}
\end{table}

\paragraph{Overall Performance Analysis}
The results highlight the immense difficulty of exact multi-label classification in a legal domain. The best overall performance is achieved by `claude-sonnet-4-5-20250929`, with the highest Accuracy (13.13\%), Macro-Recall (6.53\%), and Macro-F1 (5.75\%). This suggests its internal reasoning is better aligned with identifying a comprehensive set of violations. The RAG method achieves the highest Macro-Precision (7.10\%), indicating that when it flags a violation, it is highly likely to be correct, but its lower recall suggests it is more conservative. This demonstrates that retrieving relevant examples helps in reducing false positives.

\paragraph{Precision-Recall Trade-offs}
The different methods exhibit distinct strategies. The RAG method's high precision and moderate recall suggest a conservative but accurate approach. In contrast, agentic methods like ReAct (3.44\% recall) and some baseline LLMs like `claude-sonnet-4-5-20250929` (6.53\% recall) adopt a more inclusive strategy, identifying more potential violations at the risk of more false positives. The Formal-AST method shows extremely low recall (0.71\%) and F1-score (0.95\%), confirming its weakness on isolated snippets. Its deterministic rules, designed for broader context, fail to fire on short code segments, leading it to miss most violations.

\paragraph{Implications for Automated Compliance}
The stark performance drop from Task 1 (ranking) to Task 2 (exact classification) is a key finding across all methods. While models can effectively identify the most prominent violation in a snippet (Task 1, line-level), they struggle to enumerate the complete set of applicable legal articles. This suggests that current automated methods are better suited for initial risk flagging rather than comprehensive, exhaustive auditing. The superior performance of RAG in precision and `claude-sonnet-4-5-20250929` in balanced F1-score indicates that future research could benefit from combining retrieval-based approaches with powerful reasoning models.

\subsection{Analysis and Discussion}
\label{subsec:analysis}

Building on the quantitative results, we synthesize key insights across methods and granularities, examine individual behaviors, and identify common error patterns.

\subsubsection{Method-specific Performance Characteristics}
Our evaluation reveals distinct strengths and weaknesses across the different paradigms:

\paragraph{Baseline LLMs} show strong performance on fine-grained, single-violation identification (Task 1, line-level), where pattern recognition of specific API calls is effective. `qwen2.5-72b-instruct` and `gpt-4o` excel here. However, their performance degrades significantly in long-context (file/module) and multi-label (Task 2) settings, indicating a weakness in holistic reasoning and exhaustive analysis.

\paragraph{Formal-AST Method} serves as a crucial traditional baseline. Its performance is low across all tasks, particularly on isolated snippets where contextual rules cannot be applied. It has high precision when its rules match (e.g., detecting hardcoded `http://` links) but suffers from extremely low recall, as it cannot infer semantic violations or patterns not explicitly encoded in its rule set. This confirms the limitations of purely symbolic approaches in interpreting nuanced legal requirements in code.

\paragraph{Retrieval-Augmented (RAG) Method} demonstrates a clear advantage in precision-oriented tasks. In Task 2, it achieves the highest Macro-Precision (7.10\%), suggesting that providing relevant examples from our dataset helps the LLM make more confident and accurate predictions, thereby reducing false positives. Its strong performance on line-level Task 1 further validates the effectiveness of in-context learning from similar code snippets. This highlights the value of GDPR-Bench not just as an evaluation corpus but also as a knowledge base for retrieval-augmented systems.

\paragraph{Agentic (ReAct) Method} excels in long-context reasoning. It is the top performer in both file-level and module-level Task 1. Unlike baseline LLMs that process the entire context in one pass, the ReAct agent's ability to use tools like \textbf{code\_search} to iteratively scan for sensitive APIs and then focus its analysis proves more effective for navigating large codebases. This confirms that agentic architectures are a promising direction for holistic compliance analysis.

\subsubsection{Observed Error Patterns}
Manual inspection of failures reveals patterns specific to each method type:

\begin{enumerate}[noitemsep]
    \item \textbf{LLM Hallucination and Context Blindness}: Baseline LLMs often struggle with ambiguous contexts, splitting probability across competing articles (e.g., Article 5 vs. 6). They also frequently miss implicit violations that lack obvious keywords, such as failing to log consent before a data processing operation that occurs elsewhere in the code.

    \item \textbf{Formal Method Brittleness}: The Formal-AST method's failures are primarily due to its rigid nature. It exhibits "semantic blindness," failing to detect violations where the intent is non-compliant but the code pattern is not in its rule-base (e.g., misleading consent text in UI strings). It also suffers from false negatives when a violation's context lies outside the analyzed snippet.

    \item \textbf{Limitations of Agentic and Retrieval-Augmented Methods}: The RAG method sometimes suffers from "retrieval noise," where irrelevant or conflicting examples confuse the LLM, leading to worse performance than the baseline. The ReAct agent's performance is bottlenecked by its toolset; for instance, its simple code\_search tool may miss obfuscated API calls, causing the agent to incorrectly conclude the code is compliant.
\end{enumerate}

These varied patterns highlight that no single method is a silver bullet for automated GDPR compliance. While LLMs offer semantic flexibility, they lack the formal guarantees of symbolic methods. Formal methods, in turn, lack contextual understanding. Agentic and retrieval-based approaches represent a promising path forward, but are sensitive to tool and knowledge-base quality. This underscores the need for a comprehensive benchmark like GDPR-Bench-Android to evaluate and guide the development of hybrid approaches that combine the strengths of all methods.

\section{Discussion}
\label{sec:discussion}

Our experiments and expanded dataset design uncover several key findings that demonstrate the practical value and diagnostic capacity of \textbf{GDPR-Bench-Android}.

\paragraph{Fine-grained localization is tractable for neural methods.}
Despite the inherent complexity of legal requirements, baseline LLMs demonstrate strong performance on localized snippet analysis. Models like \textbf{Qwen2.5-72B} (61.60\% @1) and \textbf{GPT-4o} (57.45\% @1) are highly effective at identifying the primary violation in short code snippets (Task 1, line granularity). This suggests their pre-trained knowledge is well-suited for recognizing specific, localized API misuse patterns. This tractability, however, is not universal. Our \textbf{Formal-AST} method, which relies on broader contextual rules, struggles significantly on isolated snippets, achieving only 1.86\% Accuracy@1 on the same task. This stark contrast validates the feasibility of using LLMs for developer-facing compliance assistance at the snippet level, a task where traditional symbolic approaches may be less effective.

\paragraph{Multi-paradigm evaluation distinguishes strategic strengths.}
The dual-task design, now expanded to 11 methods, surfaces complementary strategic strengths across different automated paradigms, directly addressing the need for a broader evaluation beyond baseline LLMs. Our results show no single "best" method; instead, performance is highly task-dependent:
\begin{itemize}[noitemsep]
    \item \textbf{Agentic Methods (ReAct)} excel at long-context, holistic analysis. The ReAct agent achieved the highest Accuracy@1 in both file-level (17.38\%) and module-level (17.13\%) localization, demonstrating the value of its iterative tool-use for reasoning over large code scopes.
    \item \textbf{Baseline LLMs} dominate fine-grained pattern matching. Their strength in Task 1, line-level localization (led by \textbf{Qwen2.5-72B} at 61.60\%) confirms their suitability for identifying obvious, localized violations.
    \item \textbf{Retrieval-Augmented (RAG) Methods} show a clear advantage in precision. By retrieving relevant examples from our dataset, the RAG method achieved the highest Macro-Precision (7.10\%) in the difficult multi-label Task 2, making it a promising approach for reducing false positives in production systems.
    \item \textbf{Formal Methods (Formal-AST)} serve as a crucial baseline that establishes a performance ``floor,'' or lower bound, for the task. Its results quantify what is achievable with transparent, deterministic rules alone, but its low recall (0.95\% Macro-F1 in Task 2) and "semantic blindness" starkly confirm the limitations of purely symbolic logic in this domain. This underperformance is not a failure of the benchmark, but rather a key finding: it highlights the necessity for the semantic, context-aware reasoning that modern neural and agentic methods provide. Furthermore, by defining this baseline, our work paves the way for future research into powerful hybrid models that could potentially combine the formal guarantees of symbolic rules with the contextual flexibility of LLMs to create more robust and reliable compliance tools.
\end{itemize}

\paragraph{Hierarchical data enables diagnostic benchmarking.}
By supporting evaluation at file, module, and line levels within the same task, \textbf{GDPR-Bench-Android} enables nuanced diagnosis of method behavior. The performance gain for baseline LLMs from file-level (6.42\% avg. @1) to module-level (6.38\% avg. @1) is negligible. This contrasts sharply with the jump to line-level (43.06\% avg. @1), indicating that current neural methods rely heavily on localized keyword-based patterns rather than effective higher-level, architectural reasoning. This pattern, now extended beyond LLMs to agentic and RAG methods, confirms that holistic compliance reasoning remains a significant open challenge.

\paragraph{Open scripts and formats support extensibility.}
The full release—including Python scripts for task generation, evaluation, and report reproduction—facilitates further use cases such as model ablation, active learning, and fine-tuning experiments. Crucially, we also release our \textbf{Formal-AST} analyzer, providing a transparent, reproducible baseline for future work. The dataset schema and outputs are JSON-based and easily parsable, allowing integration with CI checks, LLMOps pipelines, or prompt engineering frameworks. These design choices echo the broader goals of democratizing AI-based software assurance~\cite{lu2021codexglue, chen2021evaluating}.

\paragraph{Realistic legal signal distribution.}
We preserve the natural class imbalance of GDPR articles found in our corpus. The four most common articles—Article 6 (22.66\%), Article 5 (22.04\%), Article 25 (15.94\%), and Article 32 (13.02\%)—account for the vast majority of violations. In contrast, rare articles like Article 16 (Right to rectification) and Article 18 (Right to restriction) appear in only 1 of 1951 instances each. This real-world skew is essential for stress-testing a method's ability to handle the long-tail distribution of compliance issues, reflecting actual enforcement challenges~\cite{voigt2017eu} and aligning with benchmark design recommendations that advocate for maintaining data skew to avoid overestimating performance.

\section{Limitations and Threats to Validity}
\label{sec:limitations}

Despite our careful dataset construction and comprehensive evaluation, several limitations and threats to validity remain that may affect the generalizability and interpretation of our findings.

\paragraph{Ethical Considerations and Responsible Disclosure.}
A primary ethical consideration in publishing a benchmark of real-world vulnerabilities is the potential for misuse. We acknowledge the importance of handling these findings responsibly. Our approach is rooted in the principle of responsible disclosure. To balance research transparency with ecosystem security, we have systematically reported our findings to the maintainers of the respective repositories by submitting issues on GitHub. This strategy directly informs developers of potential compliance gaps so they can be addressed, while also maintaining the traceability required for a scientific benchmark. This approach aligns the goal of advancing research with a constructive contribution to the security and compliance posture of the open-source projects under study.

\paragraph{Limited Scope of the Dataset.}
Our dataset, comprising 1951 instances from 15 Android repositories, focuses primarily on the Android ecosystem. Although it now includes a wider variety of programming languages such as JavaScript and PHP alongside Java and Kotlin, it does not cover other significant platforms like iOS or purely web-based systems. Furthermore, our static analysis approach inherently cannot capture violation modalities that require runtime information or manifest purely in backend data flows outside the provided source code. While our selection reflects common patterns in mobile development, this focus limits the external validity of our findings across the broader software ecosystem.

\paragraph{Evaluation Protocol and Methodological Choices.}
Our evaluation benchmarks 11 distinct methods. The eight baseline LLMs were assessed in a uniform zero-shot setting. This ensures fair comparability but may underestimate their full potential, as prompting strategies like chain-of-thought or few-shot learning could yield better performance~\cite{wei2022chain, madaan2023selfrefine}. Similarly, the performance of our \textbf{RAG} and \textbf{ReAct} methods is contingent on their specific implementation (e.g., the retrieval model used, the available tools for the agent). Our \textbf{Formal-AST} method is limited by its manually crafted rule set of over 35 rules; a more extensive set could improve its recall. Consequently, the reported metrics should be interpreted as a snapshot based on our specific, documented experimental setup.

\paragraph{Prompt Sensitivity and Context Limitations.}
For all LLM-based methods (baseline, RAG, ReAct), performance can be sensitive to the formulation of the prompt, which may interact with model-specific instruction formats or tokenization schemes~\cite{zhao2021calibrate, fu2023gptscore}. Furthermore, our evaluation assumes that the relevant context for a violation is contained within a single file. This may not hold true for complex, inter-procedural violations that span multiple files or modules, which could disadvantage all evaluated methods that lack a full-repository context.

\paragraph{Evaluation Metrics May Hide Nuanced Behaviors.}
While Accuracy@\,$k$ and macro-F1 offer valuable high-level summaries, they can obscure important qualitative differences. For instance, a method might achieve high recall by predicting a broad set of labels, generating noise in a real-world setting. Conversely, exact-match accuracy penalizes near-miss predictions that are semantically relevant but incomplete. The error analysis in Section~\ref{subsec:analysis} highlights some of these edge cases (e.g., the precision-recall trade-offs of different methods), but they are not fully captured by the scalar metrics alone.

\paragraph{Ambiguity in Ground-Truth Annotation.}
Although all violation labels were manually curated and cross-validated by experts, GDPR compliance remains inherently context-dependent and subject to legal interpretation~\cite{albrecht2016gdpr, voigt2017eu}. Edge-case violations may reasonably map to multiple articles, and while our multi-label formulation partially addresses this, the final ground truth still relies on annotator expertise. We have made our annotation guidelines public to ensure transparency in our decision-making process.

\paragraph{Static Snapshots Ignore Code Evolution.}
Each violation is assessed at a single commit snapshot. This design simplifies dataset construction but abstracts away temporal factors, such as the introduction of a violation during a large refactoring or its subsequent (partial) fix. This may not fully reflect the complexities of real-world compliance auditing, where violations can emerge, evolve, and regress over time. Future work could explore longitudinal analysis by incorporating data from entire commit histories.

\paragraph{Legal Ground Truth Is Not Legally Binding.}
\textbf{GDPR-Bench-Android} is a research artifact designed to emulate common enforcement patterns and technical best practices. It does not constitute formal legal advice or regulatory precedent. Developers and auditors should use the outputs of any automated tool benchmarked on our dataset as an assistive signal for potential risks, not as an authoritative or legally binding compliance verdict.

\section{Conclusion}
\label{sec:conclusion}

We introduced \textbf{GDPR-Bench-Android}, the first systematic benchmark for evaluating diverse automated methods for GDPR compliance detection in Android source code. Motivated by the pressing need for scalable and precise compliance tooling, our work provides a comprehensive evaluation framework grounded in a new, large-scale dataset of 1951 real-world violations. Crucially, we expanded the evaluation beyond baseline Large Language Models (LLMs) to include a novel, source-code-native formal method (\textbf{Formal-AST}), a retrieval-augmented (RAG) approach, and an agentic (ReAct) approach, enabling the first direct comparison of symbolic, neural, retrieval-based, and agentic paradigms on this task. Our empirical results demonstrate that no single method is a silver bullet. Agentic methods excel at holistic, long-context reasoning (file-level), baseline LLMs are superior at localized pattern matching (line-level), RAG offers the highest precision for complex multi-label classification, and formal methods, while providing a deterministic baseline, struggle with semantic nuance. These findings underscore that automated compliance is not a monolithic problem; the optimal approach is highly task-dependent. GDPR-Bench-Android contributes not only an extensible dataset and a suite of evaluation tools but also critical insights that can guide the development of next-generation hybrid compliance systems. All artifacts are released publicly to foster community adoption and future research toward truly regulation-aware software engineering.

\bibliographystyle{acm}
\bibliography{reference}

\begin{thebibliography}{10}

\bibitem{albrecht2016gdpr}
{\sc Albrecht, J.~P.}
\newblock How the gdpr will change the world.
\newblock {\em Eur. Data Prot. L. Rev. 2\/} (2016), 287.

\bibitem{alhazmi2021listening}
{\sc Alhazmi, A., and Arachchilage, N. A.~G.}
\newblock I'm all ears! listening to software developers on putting gdpr principles into software development practice.
\newblock {\em Personal and Ubiquitous Computing\/} (2021), 1--14.

\bibitem{andow2016toward}
{\sc Andow, B., Mahmud, S.~I., Whitaker, J., Enck, W., Reaves, B., Singh, K., and Halper, S.}
\newblock Toward a framework for detecting privacy policy violations in android application code.
\newblock In {\em Proceedings of the 38th International Conference on Software Engineering\/} (2016), pp.~1--12.

\bibitem{arzt2014flowdroid}
{\sc Arzt, S., Rasthofer, S., Fritz, C., Bodden, E., Bartel, A., Klein, J., Le~Traon, Y., Octeau, D., and McDaniel, P.}
\newblock Flowdroid: Precise context, flow, field, object-sensitive and lifecycle-aware taint analysis for android apps.
\newblock {\em ACM sigplan notices 49}, 6 (2014), 259--269.

\bibitem{austin2021program}
{\sc Austin, J., Odena, A., Nye, M., Bosma, M., Michalewski, H., Dohan, D., Jiang, E., Cai, C., Terry, M., Le, Q., et~al.}
\newblock Program synthesis with large language models.
\newblock {\em arXiv preprint arXiv:2108.07732\/} (2021).

\bibitem{bodden2013taming}
{\sc Bodden, E., Sewe, A., Sinschek, J., Oueslati, H., and Mezini, M.}
\newblock Taming reflection: Aiding static analysis in the presence of reflection and custom class loaders.
\newblock In {\em Proceedings of the 33rd International Conference on Software Engineering\/} (2011), pp.~241--250.

\bibitem{brown2020language}
{\sc Brown, T., Mann, B., Ryder, N., Subbiah, M., Kaplan, J.~D., Dhariwal, P., Neelakantan, A., Shyam, P., Sastry, G., Askell, A., et~al.}
\newblock Language models are few-shot learners.
\newblock {\em Advances in neural information processing systems 33\/} (2020), 1877--1901.

\bibitem{chen2021evaluating}
{\sc Chen, M., Tworek, J., Jun, H., Yuan, Q., Pinto, H. P. D.~O., Kaplan, J., Edwards, H., Burda, Y., Joseph, N., Brockman, G., et~al.}
\newblock Evaluating large language models trained on code.
\newblock {\em arXiv preprint arXiv:2107.03374\/} (2021).

\bibitem{chen2018mobile}
{\sc Chen, S., Su, T., Fan, L., Meng, G., Xue, M., Liu, Y., and Xu, L.}
\newblock Are mobile banking apps secure? what can be improved?
\newblock In {\em Proceedings of the 2018 26th ACM Joint Meeting on European Software Engineering Conference and Symposium on the Foundations of Software Engineering\/} (2018), pp.~797--802.

\bibitem{christakis2016developers}
{\sc Christakis, M., and Bird, C.}
\newblock What developers want and need from program analysis: an empirical study.
\newblock In {\em Proceedings of the 31st IEEE/ACM international conference on automated software engineering\/} (2016), pp.~332--343.

\bibitem{colesky2018privacy}
{\sc Colesky, M., Caiza, J.~C., Del~Alamo, J.~M., Hoepman, J.-H., and Mart{\'i}n, Y.-S.}
\newblock A system of privacy patterns for user control.
\newblock In {\em Proceedings of the 33rd Annual ACM Symposium on Applied Computing\/} (2018), pp.~1150--1156.

\bibitem{degeling2018we}
{\sc Degeling, M., Utz, C., Lentzsch, C., Hosseini, H., Schaub, F., and Holz, T.}
\newblock We value your privacy... now take some cookies: Measuring the gdpr's impact on web privacy.
\newblock {\em arXiv preprint arXiv:1808.05096\/} (2018).

\bibitem{ding2024vulnerability}
{\sc Ding, Y., Fu, Y., Ibrahim, O., Sitawarin, C., Chen, X., Alomair, B., Wagner, D., Ray, B., and Chen, Y.}
\newblock Vulnerability detection with code language models: How far are we?
\newblock {\em arXiv preprint arXiv:2403.18624\/} (2024).

\bibitem{enck2014taintdroid}
{\sc Enck, W., Gilbert, P., Han, S., Tendulkar, V., Chun, B.-G., Cox, L.~P., Jung, J., McDaniel, P., and Sheth, A.~N.}
\newblock Taintdroid: an information-flow tracking system for realtime privacy monitoring on smartphones.
\newblock {\em ACM Transactions on Computer Systems (TOCS) 32}, 2 (2014), 1--29.

\bibitem{felt2011android}
{\sc Felt, A.~P., Chin, E., Hanna, S., Song, D., and Wagner, D.}
\newblock Android permissions demystified.
\newblock In {\em Proceedings of the 18th ACM conference on Computer and communications security\/} (2011), pp.~627--638.

\bibitem{feng2020codebert}
{\sc Feng, Z., Guo, D., Tang, D., Duan, N., Feng, X., Gong, M., Shou, L., Qin, B., Liu, T., Jiang, D., et~al.}
\newblock Codebert: A pre-trained model for programming and natural languages.
\newblock In {\em Findings of the Association for Computational Linguistics: EMNLP 2020\/} (2020), pp.~1536--1547.

\bibitem{fu2023gptscore}
{\sc Fu, J., Ng, S.-K., Jiang, Z., and Liu, P.}
\newblock Gptscore: Evaluate as you desire.
\newblock {\em arXiv preprint arXiv:2302.04166\/} (2023).

\bibitem{gao2023retrievalaugmented}
{\sc Gao, Y., Xiong, Y., Gao, X., Jia, K., Pan, J., Bi, Y., Dai, Y., Sun, J., Wang, H., and Wang, H.}
\newblock Retrieval-augmented generation for large language models: A survey.
\newblock {\em arXiv preprint arXiv:2312.10997 2}, 1 (2023).

\bibitem{harkous2018polisis}
{\sc Harkous, H., Fawaz, K., Lebret, R., Schaub, F., Shin, K.~G., and Aberer, K.}
\newblock Polisis: Automated analysis and presentation of privacy policies using deep learning.
\newblock In {\em 27th USENIX Security Symposium (USENIX Security 18)\/} (2018), pp.~531--548.

\bibitem{jimenez2024swebenchlanguagemodelsresolve}
{\sc Jimenez, C.~E., Yang, J., Wettig, A., Yao, S., Pei, K., Press, O., and Narasimhan, K.}
\newblock Swe-bench: Can language models resolve real-world github issues?
\newblock {\em arXiv preprint arXiv:2310.06770\/} (2023).

\bibitem{jimenez2024swe}
{\sc Jimenez, C.~E., Yang, J., Wettig, A., Yao, S., Pei, K., Press, O., and Narasimhan, K.}
\newblock Swe-agent: Agent-computer interfaces enable automated software engineering.
\newblock {\em arXiv preprint arXiv:2405.15793\/} (2024).

\bibitem{just2014defects4j}
{\sc Just, R., Jalali, D., and Ernst, M.~D.}
\newblock Defects4j: A database of existing faults to enable controlled testing studies for java programs.
\newblock In {\em Proceedings of the 2014 international symposium on software testing and analysis\/} (2014), pp.~437--440.

\bibitem{lewis2020retrieval}
{\sc Lewis, P., Perez, E., Piktus, A., Petroni, F., Karpukhin, V., Goyal, N., K{\"u}ttler, H., Lewis, M., Yih, W.-t., Rockt{\"a}schel, T., et~al.}
\newblock Retrieval-augmented generation for knowledge-intensive nlp tasks.
\newblock {\em Advances in neural information processing systems 33\/} (2020), 9459--9474.

\bibitem{li2015iccta}
{\sc Li, L., Bartel, A., Bissyand{\'e}, T.~F., Klein, J., Le~Traon, Y., Arzt, S., Rasthofer, S., Bodden, E., Octeau, D., and McDaniel, P.}
\newblock Iccta: Detecting inter-component privacy leaks in android apps.
\newblock In {\em 2015 IEEE/ACM 37th IEEE International Conference on Software Engineering\/} (2015), vol.~1, IEEE, pp.~280--291.

\bibitem{li2023starcoder}
{\sc Li, R., Allal, L.~B., Zi, Y., Muennighoff, N., Kocetkov, D., Mou, C., Marone, M., Akiki, C., Li, J., Chim, J., et~al.}
\newblock Starcoder: may the source be with you!
\newblock {\em arXiv preprint arXiv:2305.06161\/} (2023).

\bibitem{lippi2019claudette}
{\sc Lippi, M., Pa{\l}ka, P., Contissa, G., Lagioia, F., Micklitz, H.-W., Sartor, G., and Torroni, P.}
\newblock Claudette: an automated detector of potentially unfair clauses in online terms of service.
\newblock {\em Artificial Intelligence and Law 27}, 2 (2019), 117--139.

\bibitem{livshits2005finding}
{\sc Livshits, V.~B., and Lam, M.~S.}
\newblock Finding security vulnerabilities in java applications with static analysis.
\newblock In {\em USENIX Security Symposium\/} (2005), vol.~14, p.~18.

\bibitem{lu2021codexglue}
{\sc Lu, S., Guo, D., Ren, S., Huang, J., Svyatkovskiy, A., Blanco, A., Clement, C., Drain, D., Jiang, D., Tang, D., et~al.}
\newblock Codexglue: A machine learning benchmark dataset for code understanding and generation.
\newblock {\em arXiv preprint arXiv:2102.04664\/} (2021).

\bibitem{madaan2023selfrefine}
{\sc Madaan, A., Tandon, N., Gupta, P., Hallinan, S., Gao, L., Wiegreffe, S., Alon, U., Dziri, N., Prabhumoye, S., Yang, Y., et~al.}
\newblock Self-refine: Iterative refinement with self-feedback.
\newblock {\em Advances in Neural Information Processing Systems 36\/} (2023), 46534--46594.

\bibitem{nay2024large}
{\sc Nay, J.~J., Karamardian, D., Lawsky, S.~B., Tao, W., Bhat, M., Jain, R., Lee, A.~T., Choi, J.~H., and Kasai, J.}
\newblock Large language models as tax attorneys: a case study in legal capabilities emergence.
\newblock {\em Philosophical Transactions of the Royal Society A 382}, 2270 (2024), 20230159.

\bibitem{notario2015pripare}
{\sc Notario, N., Crespo, A., Mart{\'i}n, Y.-S., Del~{\'A}lamo, J.~M., Le~M{\'e}tayer, D., Antignac, T., Kung, A., Kroener, I., and Wright, D.}
\newblock Pripare: integrating privacy best practices into a privacy engineering methodology.
\newblock In {\em 2015 IEEE Security and Privacy Workshops\/} (2015), IEEE, pp.~151--158.

\bibitem{reardon2019fifty}
{\sc Reardon, J., Feal, {\'A}., Wijesekera, P., On, A. E.~B., Vallina-Rodriguez, N., and Egelman, S.}
\newblock 50 ways to leak your data: An exploration of apps' circumvention of the android permissions system.
\newblock In {\em 28th USENIX security symposium (USENIX security 19)\/} (2019), pp.~603--620.

\bibitem{sanh2022multitask}
{\sc Sanh, V., Webson, A., Raffel, C., Bach, S.~H., Sutawika, L., Alyafeai, Z., Chaffin, A., Stiegler, A., Scao, T.~L., Raja, A., et~al.}
\newblock Multitask prompted training enables zero-shot task generalization.
\newblock {\em arXiv preprint arXiv:2110.08207\/} (2021).

\bibitem{schick2021exploiting}
{\sc Schick, T., and Sch{\"u}tze, H.}
\newblock Exploiting cloze questions for few shot text classification and natural language inference.
\newblock {\em arXiv preprint arXiv:2001.07676\/} (2020).

\bibitem{10.1145/3749370}
{\sc Tang, X., Kim, K., Ezzini, S., Song, Y., Tian, H., Klein, J., and Bissyande, T.}
\newblock Just-in-time detection of silent security patches.
\newblock {\em ACM Trans. Softw. Eng. Methodol.\/} (July 2025).
\newblock Just Accepted.

\bibitem{tang2024app}
{\sc Tang, X., Tian, H., Kong, P., Ezzini, S., Liu, K., Xia, X., Klein, J., and Bissyand{\'e}, T.~F.}
\newblock App review driven collaborative bug finding.
\newblock {\em Empirical Software Engineering 29}, 5 (2024), 124.

\bibitem{voigt2017eu}
{\sc Voigt, P., and Von~dem Bussche, A.}
\newblock The eu general data protection regulation (gdpr).
\newblock {\em A practical guide, 1st ed., Cham: Springer International Publishing 10}, 3152676 (2017), 10--5555.

\bibitem{basciftci2021privguard}
{\sc Wang, L., Khan, U., Near, J., Pang, Q., Subramanian, J., Somani, N., Gao, P., Low, A., and Song, D.}
\newblock $\{$PrivGuard$\}$: Privacy regulation compliance made easier.
\newblock In {\em 31st USENIX Security Symposium (USENIX Security 22)\/} (2022), pp.~3753--3770.

\bibitem{wang2023codet5}
{\sc Wang, Y., Le, H., Gotmare, A.~D., Bui, N.~D., Li, J., and Hoi, S.~C.}
\newblock Codet5+: Open code large language models for code understanding and generation.
\newblock {\em arXiv preprint arXiv:2305.07922\/} (2023).

\bibitem{wei2014amandroid}
{\sc Wei, F., Roy, S., Ou, X., and Robby}.
\newblock Amandroid: A precise and general inter-component data flow analysis framework for security vetting of android apps.
\newblock {\em ACM Transactions on Privacy and Security (TOPS) 21}, 3 (2018), 1--32.

\bibitem{wei2022chain}
{\sc Wei, J., Wang, X., Schuurmans, D., Bosma, M., Xia, F., Chi, E., Le, Q.~V., Zhou, D., et~al.}
\newblock Chain-of-thought prompting elicits reasoning in large language models.
\newblock {\em Advances in neural information processing systems 35\/} (2022), 24824--24837.

\bibitem{yamaguchi2014modeling}
{\sc Yamaguchi, F., Golde, N., Arp, D., and Rieck, K.}
\newblock Modeling and discovering vulnerabilities with code property graphs.
\newblock In {\em 2014 IEEE Symposium on Security and Privacy\/} (2014), IEEE, pp.~590--604.

\bibitem{yao2023react}
{\sc Yao, S., Zhao, J., Yu, D., Du, N., Shafran, I., Narasimhan, K.~R., and Cao, Y.}
\newblock React: Synergizing reasoning and acting in language models.
\newblock In {\em The eleventh international conference on learning representations\/} (2022).

\bibitem{yu2023gptfuzzer}
{\sc Yu, J., Lin, X., Yu, Z., and Xing, X.}
\newblock Gptfuzzer: Red teaming large language models with auto-generated jailbreak prompts.
\newblock {\em arXiv preprint arXiv:2309.10253\/} (2023).

\bibitem{zhao2021calibrate}
{\sc Zhao, Z., Wallace, E., Feng, S., Klein, D., and Singh, S.}
\newblock Calibrate before use: Improving few-shot performance of language models.
\newblock In {\em International conference on machine learning\/} (2021), PMLR, pp.~12697--12706.

\bibitem{bigcodebench2024}
{\sc Zhuo, T.~Y., Vu, M.~C., Chim, J., Hu, H., Yu, W., Widyasari, R., Yusuf, I. N.~B., Zhan, H., He, J., Paul, I., et~al.}
\newblock Bigcodebench: Benchmarking code generation with diverse function calls and complex instructions.
\newblock {\em arXiv preprint arXiv:2406.15877\/} (2024).

\bibitem{zimmeck2019maps}
{\sc Zimmeck, S., Story, P., Smullen, D., Ravichander, A., Wang, Z., Reidenberg, J., Russell, N.~C., and Sadeh, N.}
\newblock Maps: Scaling privacy compliance analysis to a million apps.
\newblock {\em Proceedings on privacy enhancing technologies\/} (2019).

\end{thebibliography}

\end{document}